\documentclass[fleqn,usenatbib]{mnras}
\usepackage{newtxtext,newtxmath}
\usepackage[utf8]{inputenc}
\usepackage[T1]{fontenc}
\usepackage{booktabs}
\usepackage{threeparttable}
\usepackage{array, caption, threeparttable}
\usepackage[font=small,labelfont=bf,labelsep=none]{caption}
\usepackage{float}
\usepackage{graphicx}
\usepackage{float}
\usepackage{subfig}
\usepackage{xcolor}
\usepackage{tabularx}
\usepackage[T1]{fontenc}
\DeclareRobustCommand{\VAN}[3]{#2}
\let\VANthebibliography\thebibliography
\def\thebibliography{\DeclareRobustCommand{\VAN}[3]{##3}\VANthebibliography}
\usepackage{graphicx}	
\usepackage{amsmath}	

\title[Typing SNR G352.7$-$0.1]{
Typing supernova remnant G352.7$-$0.1 using XMM-Newton X-ray observations
}

\author[Ling-Xiao Dang et al.]{
Ling-Xiao Dang, $^{1}$
Ping Zhou, $^{1,2}$ \thanks{E-mail:pingzhou@nju.edu.cn}
Lei Sun, $^{1}$
Junjie Mao, $^{3}$
Jacco Vink, $^{4}$
Qian-Qian Zhang, $^{1}$
\newauthor
and Vladimír Domček $^{4}$
\\
$^{1}$School of Astronomy $\&$ Space Science, Nanjing University, 163 Xianlin Avenue, Nanjing 210023, People's Republic of China \\
$^{2}$ Key Laboratory of Modern Astronomy and Astrophysics, Nanjing University, Ministry of Education, Nanjing 210023, People's Republic of China \\
$^{3}$ Department of Astronomy, Tsinghua University, Haidian DS 100084, Beijing, People's Republic of China\\
$^{4}$ Anton Pannekoek Institute for Astronomy, University of Amsterdam, Science Park 904, 1098 XH Amsterdam, The Netherlands\\
}

\date{Accepted XXX. Received YYY; in original form ZZZ}
\pubyear{2015}
\begin{document}
\label{firstpage}
\pagerange{\pageref{firstpage}--\pageref{lastpage}}
\maketitle
\begin{abstract}

G352.7$-$0.1 is a mixed-morphology (MM) supernova remnant (SNR) with multiple radio arcs and has a disputed supernova origin. We conducted a spatially resolved spectroscopic study of the remnant with XMM-Newton X-ray data to investigate its explosion mechanism and explain its morphology.
The global X-ray spectra of the SNR can be adequately reproduced using a metal-rich thermal plasma model with a temperature of $\sim 2$~keV and ionization timescale of $\sim 3\times 10^{10}~{\rm cm^{-3}~s}$. 
Through a comparison with various supernova nucleosynthesis models, we found that observed metal properties from Mg to Fe can be better described using core-collapse supernova models, while thermonuclear models fail to explain the observed high Mg/Si ratio.
The best-fit supernova model suggests a $\sim 13$~$M_\odot$ progenitor star, consistent with previous estimates using the wind bubble size.
We also discussed the possible mechanisms that may lead to SNR G352.7$-$0.1 being an MMSNR. 
By dividing the SNR into several regions, we found that the temperature and abundance do not significantly vary with regions, except for a decreased temperature and abundance in a region interacting with molecular clouds. 
The brightest X-ray emission of the SNR spatially matches with the inner radio structure, suggesting that the centrally filled X-ray morphology results from a projection effect.  
\end{abstract}


\section{Introduction}
A supernova remnant (SNR) is formed from the interaction between a supernova (SN) and its surrounding environment. 
It contains information about the environment where the SN occurred and the SN explosion mechanism itself.
A few methods have been proposed to 
infer the explosion mechanisms that created the SNRs based on X-ray observations.
Statistically, core-collapse (CC) SNRs exhibit a higher degree of asymmetry than Type \uppercase\expandafter{\romannumeral1}a SNRs \citep{Lopez2009, Lopez2011}.
In addition to the hint from the SNR morphology, the X-ray spectral analysis of SNR metals can also provide clues to the SN explosion mechanism, as different metal yields are predicted for different SNe \citep[see references therein]{Vink2020}.
\cite{Hughes1995} showed that the metal abundances obtained from X-ray spectra could be used to distinguish Type Ia and CC SNRs because Type Ia SNR spectra tend to show stronger emission from Si to Fe. 
Another method to discriminate between progenitor types is using the Fe K$\alpha$ emission line, which tends to appear below 6.55 keV for Type \uppercase\expandafter{\romannumeral1}a SNRs \citep{Yamaguchi2014}.

The traditional classification of SNRs is mainly based on the SNR morphology in the radio and X-ray band, rather than the SN type.
There are four types of SNRs: shell-type, filled-center, composite, and mixed-morphology (or thermal composite). Mixed-morphology supernova remnants (MMSNRs) are a class with a filled-center morphology in X-rays and a shell-like shape in the radio band \citep{Rho1998ApJ...503L.167R, Jones1998PASP..110.1059J}. Initially, it was thought that the thermal emission in the interior of MMSNRs came from low-abundance hot gas, but later observations have revealed enhanced abundances in some MMSNRs \citep{Lazendic2006ApJ...647..350L, Bocchino_2009}.  Although the origin of the different X-ray and radio morphologies is still unclear, an interesting phenomenon is that most MMSNRs are interacting with adjacent molecular or H\uppercase\expandafter{\romannumeral1} clouds \citep{Rho1998ApJ...503L.167R}.
Due to the association between massive stars and molecular clouds, MMSNRs are sometimes thought to be remnants of CCSNe \citep{Rho1998ApJ...503L.167R}. 
The presence of associated pulsars in some MMSNRs \citep[e.g., W44, IC 443,][]{Wolszczan1991ApJ...372L..99W, Olbert2001ApJ...554L.205O} has supported this opinion.
However, a few MMSNRs \citep[e.g., G344.7$-$0.1, W49B, Sgr~A~East,][]{Yamaguchi_2012,Zhou2018,zhousgr2021} are proposed to have a Type \uppercase\expandafter{\romannumeral1}a origin, disfavoring the opinion that the morphology can be regarded as a good indicator of SN type.

G352.7$-$0.1 has been classified as an MMSNR based on its X-ray and radio morphology \citep{Giacani2009}.
This SNR was first discovered in the radio band by \cite{Clark1973}. Subsequent radio observations revealed a biannular radio morphology and gave a rough distance of 11~kpc \citep{Dubner1993} using the radio surface brightness -- diameter ($\Sigma$--$D$) relationship \citep{Huang1985}. Recently, \cite{Zhang2023} updated the SNR distance to 10.5~kpc by studying the interaction between the SNR and the molecular cloud. 
In the X-ray band, the SNR reveals clumpy structures and diffuse emissions filling the interior. The X-ray spectra can be fit with one or two non-equilibrium ionization (NEI) plasma models with over-solar abundances, suggesting that the X-ray emission is ejecta-dominated \citep{Kinugasa1998,Giacani2009,Pannuti2014,Sezer2014}.

There is still no consensus on the formation of the MM for G352.7$-$0.1. \cite{Giacani2009} proposed that the barrel-shaped structure observed in the radio band results from the SN shock expanding within the axially symmetric stellar wind of a massive progenitor star. The centrally-filled X-ray morphology could be attributed to enhanced emission due to metal overabundance, coupled with increased thermal conduction effects in the SNR interior \citep{Pannuti2014}. 
Using three-dimensional (3D) hydrodynamical simulations, \cite{Toledo2014} found that an SN exploding near the border of a dense cloud can create the multiple-ring radio structure and centrally filled X-ray morphology.

\begin{figure*}
 	\centering
	\includegraphics[width=0.8\textwidth]{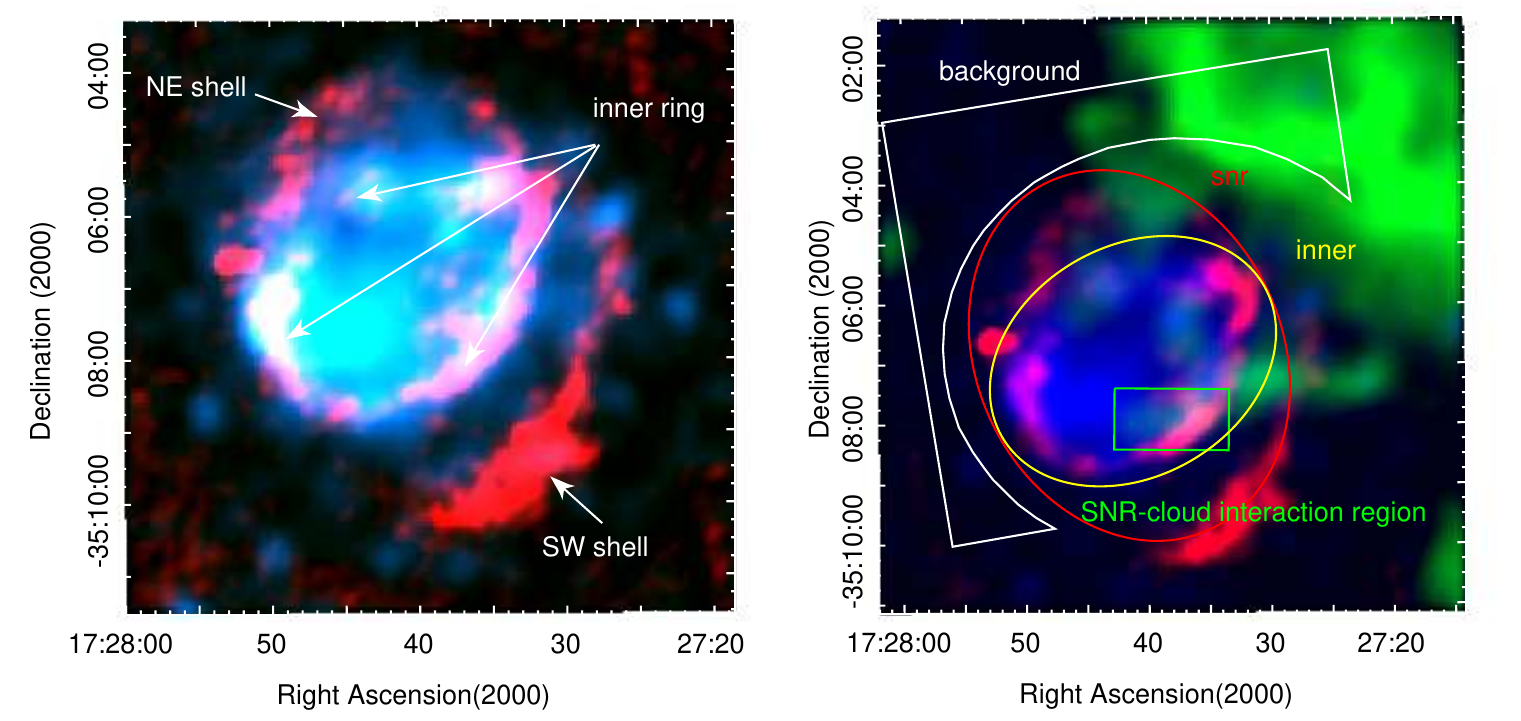}
    \caption{Composite image of SNR G352.7$-$0.1. Left image: Red: VLA 4.8 GHz radio continuum image \citep{Giacani2009}; Cyan: XMM-Newton 0.8--7 keV X-ray image. Right image: Red: VLA 4.8 GHz radio continuum image \citep{Giacani2009}; Green: APEX $^{12}$CO emission image \citep{Zhang2023}; Blue: XMM-Newton 0.8--7 keV X-ray image. The red, yellow, and white, regions denote the spectral extraction regions for the global SNR plasma (``snr''), the X-ray-bright interior gas (``inner''), and the background, respectively. The region where the SNR interacts with molecular cloud \citep{Zhang2023} is marked with a green box.
    }
    \label{fig:rgb}
\end{figure*}

Whether SNR G352.7$-$0.1 resulted from a Type-Ia or CC SN explosion is not clear.
\cite{Giacani2009} suggested that the progenitor could be a massive star, considering that the barrel-shape morphology may be shaped by the axially symmetric wind of a red supergiant.
The CC origin of the SNR is also supported by \cite{Pannuti2014}, who claimed that the massive progenitor star can explain the large swept-up mass of $45~\rm{M_\odot}$ coupled with the X-ray-emitting ejecta mass of 2.6 $\rm{M_\odot}$.
In contrast, a Type \uppercase\expandafter{\romannumeral1}a origin for SNR G352.7$-$0.1 is proposed due to the low centroid energy of the Fe K$\alpha$ \citep{Yamaguchi2014,Sezer2014,Fujishige2023}. 
Assuming a Type Ia origin and considering the SNR morphology and massive swept-up gas, \cite{soker2024} suggest that G352.7$-$0.1 involves a peculiar Type Ia SN inside a planetary nebula.

Considering that the X-ray emission of this SNR is dominated by ejecta, it is worthwhile to study the metal pattern of the SNR to infer the SN explosion mechanism and progenitor star.
In this paper, we revisited the X-ray data of G352.7$-$0.1, aiming to disentangle the disputes of its SN type and morphology. 
Section \ref{sec:2} describes the XMM-Newton data and data reduction. Section \ref{sec:3} presents the multi-band images and spatially resolved spectral analysis of XMM-Newton data. 
In Section \ref{sec:discussion}, we discussed the plasma properties and the evolution parameters of the SNR. We further investigate the explosion mechanism and progenitor star, and provide interpretations for the morphology of G352.7$-$0.1. 
Finally, the conclusions are summarized in Section \ref{sec:5}. 

\section{Data}\label{sec:2}

SNR G352.7$-$0.1 was observed using the European Photon Imaging Camera (EPIC) on board the XMM-Newton X-ray telescope (Obs.\ ID: 0150220101; PI: J. Hughes) on October 3, 2002.
The EPIC cameras operate within an energy range of 0.15 to 15 keV and provide a field of view (FOV) of $30'$ and an angular resolution of $\sim6''$ at 1 keV.
During this observation, the XMM-Newton telescope was operated in MEDIUM filter mode, and the pn camera was in Full Frame mode.  
The exposure time was 30.5, 30.5, and 19.9 ks for the MOS1, MOS2, and pn, respectively.
After removing the time intervals with proton flare contamination, the net exposure time used in our analysis is around 29.3, 29.8, and 17.0 ks for the MOS1, MOS2, and pn, respectively.

We used the HEASoft software (version 6.30.1) and the Science Analysis System (SAS) package (Version 20.0.0) for data reduction and Xspec (Version 12.12.1)  for spectral analysis.

For comparison purposes, we also used $^{12}$CO~$J=2$--1 molecular line image observed with the Atacama Pathfinder Experiment (APEX) telescope \citep{Zhang2023} and the 4.8~GHz continuum image \citep{Giacani2009} observed with the Very Large Array (VLA).

\section{Results}
\label{sec:3}

\subsection{Multi-band images} 

The left panel of Figure \ref{fig:rgb} compares the radio and X-ray morphology of SNR G352.7$-$0.1. 
The radio emission of this SNR shows a double shell structure (labeled as the NE and SW shells in Figure \ref{fig:rgb}) with an inner ring.
Unlike radio morphology, X-ray emission is clumpy and mainly distributed within the inner ring, with a dim X-ray halo distributed between the inner radio ring and outer shells. 
The right panel of Figure \ref{fig:rgb} adds the distribution of the molecular gas for comparison. 
We also defined a few regions in this figure for further detailed spectral analysis.

\begin{figure}	\includegraphics[width=\columnwidth]{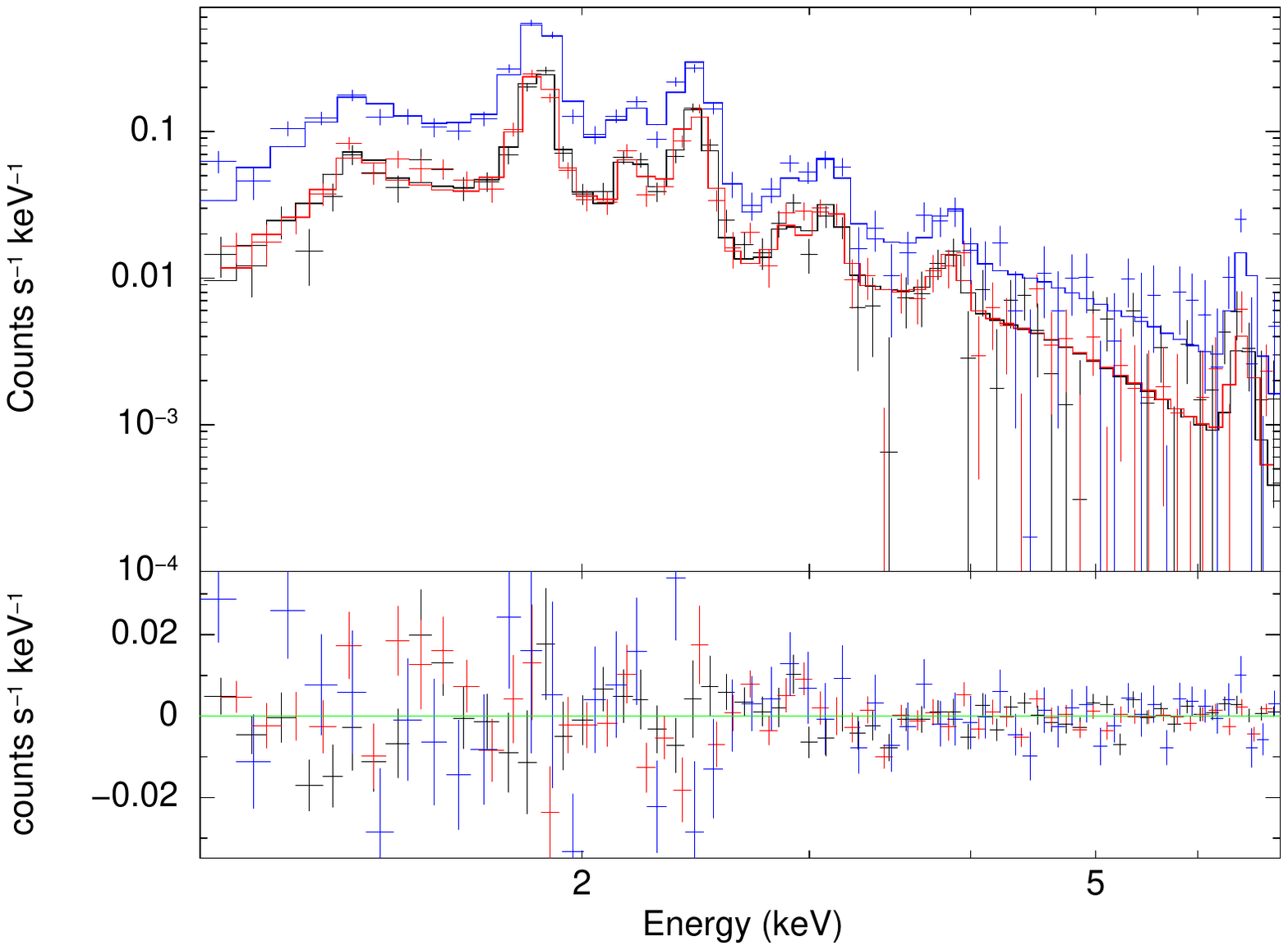}
    \caption{XMM-Newton MOS1, MOS2, and pn spectra of regions ``snr'' in the 0.8--7 keV band (black, red, and blue, respectively), fitted with absorbed $tbabs \times vnei$ models. The region selection is shown in Figure~\ref{fig:rgb}, and the best-fit results are listed in Table~\ref{tab:regions}. }
    \label{snrincludehalo}
\end{figure}

\subsection{Global spectra}

The global spectra of the whole SNR were extracted from the elliptic region ``snr'' (red ellipse) denoted in the right panel of Figure \ref{fig:rgb}. This region includes the X-ray-bright inner region (``inner''; yellow ellipse) inside the inner radio shell and two faint lobes in the northeast and southwest. 
The background spectra were selected inside a box region covering the central CCD of the MOS cameras and outside the elliptical region covering the SNR.

We used an NEI model $vnei$ to fit the spectra of G352.7$-$0.1, as suggested by previous X-ray studies \citep{Kinugasa1998, Giacani2009,Pannuti2014}.  
This plasma model assumes a constant temperature and single ionization timescale. The absorption model $tbabs$ accounts for interstellar absorption due to the atomic, grain, and molecular phases in the ISM \citep{Wilms2000}. 
We adopted solar abundances of \cite{Asplund2009}, allowed the abundances of Mg, Si, S, Ar, Ca, and Fe to vary, and tied the abundance of Ni to Fe. Other abundances cannot be constrained by the fit and thus are fixed to the solar values. We binned the data with the optimal binning method \citep{Kaastra2016} and fit the data using the C-statistics \citep{2017Kaastra}.

As shown in Figure~\ref{snrincludehalo}, the single-temperature $vnei$ model can describe the spectra from the whole SNR, with the fit results summarized in Table \ref{tab:regions}. 
We found that the collisional ionization equilibrium models, such as $vapec$, cannot reproduce the spectra. Since G352.7$-0.1$ is classified as a mixed-morphology SNR and a few SNRs in this class show recombining plasma \citep[e.g., W49B and IC443][]{Kawasaki_2005}, we also tested the recombining plasma model $vrnei$ and set an electron temperature $kT$ smaller than the initial temperature $kT_{\rm init}$. This recombining model cannot reproduce the spectra and the C-statistic is worse than that from the $vnei$ model. 
We further checked if the double-temperature plasma models can improve the spectra fit. 
However, adding a second thermal component does not significantly improve the spectral fit.
We also analyzed the spectrum of the bright inner region and obtained similar spectral parameters to that of ``snr''. This is because the X-ray emission of the SNR is dominated by the inner region.

The spectral fit suggests that the X-ray emission from G352.7$-0.1$ is best characterized by an under-ionized plasma with 
a temperature of $kT\sim 2$~keV and an ionization timescale $\tau\sim 3\times 10^{10}$~cm$^{-3}$~s. 
The X-ray-emitting plasma of the SNR is metal-rich, with abundances 
Mg=$2.8^{+2.3}_{-1.1}$, Si=$5.6^{+2.9}_{-1.4}$, S=$7.2^{+3.1}_{-1.4}$, Ar=$7.4^{+4.0}_{-2.3}$, Ca=$11.0^{+7.8}_{-4.9}$, and Fe=$10.3^{+13.2}_{-5.7}$ (Confidence levels here are 90$\%$. Unless otherwise specified in this paper, the confidence level for the error range is set at 90$\%$). 

Hereafter we compared our models and results with earlier X-ray studies of SNR G352.7$-0.1$.
Unlike previous studies that used the absorption model $phabs$, we chose $tbabs$ as the absorption model, which additionally considers the effects of the grain phase and the molecules in the interstellar medium. The solar element abundance model we used is the result obtained by \cite{Asplund2009}, which differs from the earlier solar element abundance model \citep{1989Anders} used in previous studies \citep{Sezer2014, Giacani2009}. 
Our best-fit plasma temperature ($\rm{2.1^{+0.7}_{-0.2}\ keV}$) is consistent with that obtained by \cite{Giacani2009}, but \cite{Pannuti2014}  found a lower temperature of $1.20^{+0.24}_{-0.28}$~keV using the same XMM-Newton data.
Our ionization timescale $\tau=3.0^{+0.6}_{-0.5}\times 10^{10}\rm {\ cm^{-3}\ s}$ agrees with that from  \cite{Pannuti2014} ($4.07^{+2.53}_{-1.17} \times 10^{10} \rm {\ cm^{-3}\ s}$), but is slightly lower than that obtained by \cite{Giacani2009} (4.5 $\pm$ 0.5$ \ \times 10^{10}  \rm {\ cm^{-3} \ s}$). 
In the studies by \cite{Pannuti2014} and \cite{Giacani2009}, the abundances of Mg, Ca, and Fe are fixed to the solar values. 
By using the optimal binning method as described by \cite{Kaastra2016}, we can discern the Fe and Ca lines and obtain super-solar abundances for Mg, Ca, and Fe (Ni fixed to Fe). 
Consequently, we obtained larger metal abundances than those reported by these two studies.
However, the abundances of S--Fe elements obtained in this paper are consistent with those from \cite{Sezer2014}, who used a double component model to fit the Suzaku data, but the temperature and ionization timescale values are different.

\subsection{Spatially resolved spectral analysis}
\begin{figure}	\includegraphics[width=\columnwidth]{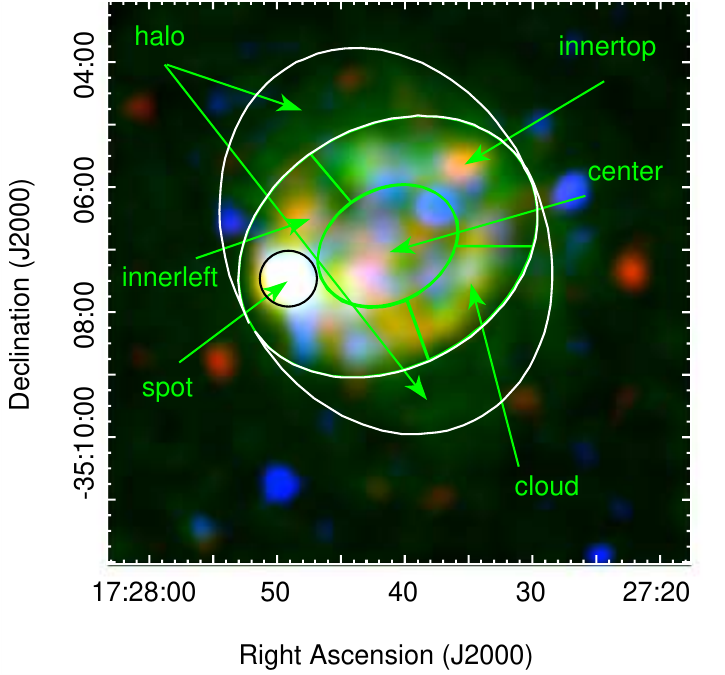}
    \caption{XMM-Newton X-ray image of G352.7$-$0.1 (Red: 0.8-1.5 keV; Green: 1.5-3.3 keV; Blue: 3.3-7 keV). The image is exposure-corrected and adaptively smoothed. The ``innerleft'' region does not include the ``spot'' region.}
    \label{fig:xrays}
\end{figure}

To search for spatial variations in plasma properties, we conducted a spatially resolved X-ray spectroscopic study of G352.7$-$0.1.  Based on the radio morphology and the X-ray structures, we divided this SNR into six regions, as shown in Figure~\ref{fig:xrays}.
In addition to the bright central areas, there are two halos on the northeastern and southwestern sides. We hypothesized an explosion scenario in which the SN occurred in a site with enhanced density. As a result, the shocks blew out in the northeast and southwest, while the expansion was slowed in other directions due to interaction with the denser medium \citep[see also][]{Zhang2023}. 
Therefore, we assume that the halos in the northeast and southwest have a similar nature. We combined the two halo regions in the spectral analysis to increase the statistics.

We used the same background region and model ($tbabs$ ${\times}$ $vnei$) as in the global spectrum.
Since the abundances and foreground column density ($N\mathrm{_{H}}$) in regions ``spot'' and ``halo'' cannot be well constrained, we 
fixed the $N\mathrm{_{H}}$ in these two regions to that of the whole SNR ($4.66\times 10^{22}$~cm$^{-2}$).
Moreover, for the ``spot'', ``halo'', and ``cloud'' regions,  we fixed the Mg abundance to 1 since it cannot be constrained, but freezing or thawing the Mg abundance does not much change the Si--Fe abundances. Table \ref{tab:regions} summarizes the fitting results.

In the ``cloud'' region, the SNR shock is impacting the molecular cloud \citep[][see also Figure~\ref{fig:rgb}]{Zhang2023}. 
We found a low temperature and metal abundance in this region, providing strong evidence for the ejecta-ISM mixing, while
the physical and chemical properties in other regions do not vary significantly across the SNR.

\subsection{Abundance ratios} \label{sec:mcmc}

The abundance ratios in SNRs can be used to infer the explosion mechanisms and progenitor systems. We calculated the element abundance ratios relative to Si,  as the reference element Si has a smaller error than heavier elements. 

Considering that the fitted abundances of different elements may not be independent of each other, we did not use the error propagation method to calculate the abundance ratios and their uncertainties. Instead, we applied the Markov Chain Monte Carlo (MCMC) method built in Xspec to provide the uncertainties of the abundance ratios.   To obtain converged results, we set the upper limit of $kT$ to 10~keV and fix $N_{\rm H}$ within the 90\% error range as obtained from the previous fit.
We applied the Goodman-Weare algorithm with 20 walkers, chain steps of $10^5$, and burn-in steps of $5\times 10^4$ to ensure the chain convergence. To generate the initial walkers, a Gaussian distribution and the covariance matrix of the fit are used. 
Using the MCMC method, we obtained the 90\% confidence range of the abundance ratios as: Mg/Si=0.43--0.66, S/Si=1.11--1.35, Ar/Si=0.97--1.65, Ca/S=1.33--2.84, Fe/Si=1.54--2.94. 

\begin{table*}
        \renewcommand\arraystretch{1.5}
        \renewcommand{\baselinestretch}{3}
	\centering
        \setlength{\abovecaptionskip}{0pt}
        \setlength{\belowcaptionskip}{10pt}
        \caption{Best-fit results and 90$\%$ uncertainties for small regions in G352.7$-$0.1}\label{tab:aStrangeTable}
	\label{tab:regions}
	\begin{tabular}{lccccccc}
	\hline\hline\noalign{\smallskip}	
	Parameter & snr & center & innerleft & innertop & cloud & spot & halo\\
	\noalign{\smallskip}\hline\noalign{\smallskip}
        $\rm {N_H}$ ($10^{22}$  ${\rm cm}^{-2}$) & $4.66^{+0.65}_{-0.73}$ & $4.48^{+0.82}_{-0.69}$ & $4.61^{+1.04}_{-0.41}$ & $3.85^{+0.96}_{-0.63}$ & $4.32^{+0.76}_{-0.98}$ & $4.66$(fixed) & $4.66$(fixed)\\
        $kT$(keV) & $2.06^{+0.68}_{-0.21}$ & $2.33^{+1.13}_{-0.66}$ & $1.76^{+0.56}_{-0.44}$ & $3.77^{+2.21}_{-1.39}$ & $0.99^{+0.56}_{-0.28}$ & $1.77^{+0.36}_{-0.30}$ & $2.50^{+1.37}_{-1.00}$\\
        Mg & $2.79^{+2.34}_{-1.11}$ & $2.43^{+2.67}_{-1.02}$ & $2.64^{+7.15}_{-1.34}$ & $2.25^{+2.15}_{-0.95}$ & $1.00$(fixed) & $1.00$(fixed) & $1.80^{+2.55}_{-1.46}$\\
        Si & $5.64^{+2.94}_{-1.43}$ & $5.23^{+4.44}_{-1.42}$ & $4.46^{+7.54}_{-1.65}$ & $6.58^{+2.83}_{-1.68}$ & $2.04^{+0.69}_{-0.50}$ & $6.20^{+6.11}_{-2.46}$ & $5.87^{+5.31}_{-2.59}$\\
        S & $7.17^{+3.05}_{-1.44}$ & $7.01^{+4.73}_{-1.72}$ & $6.22^{+8.22}_{-1.86}$ & $8.67^{+3.18}_{-2.03}$ & $2.96^{+1.19}_{-0.80}$ & $7.67^{+7.37}_{-2.93}$ & $7.67^{+6.45}_{-3.04}$ \\
        Ar & $7.42^{+3.95}_{-2.32}$ & $4.89^{+4.54}_{-2.81}$ & $7.38^{+10.82}_{-3.42}$ & $8.36^{+5.54}_{-3.81}$ & $5.94^{+5.62}_{-3.37}$ & $11.69^{+13.84}_{-6.26}$ & $6.47^{+9.90}_{-5.52}$ \\
        Ca & $10.99^{+7.81}_{-4.93}$ & $10.26^{+11.36}_{-5.41}$ & $7.28^{+15.49}_{-6.03}$ & $3.22^{+7.07}_{-3.23}$ & $17.60^{+29.96}_{-12.05}$ & $13.93^{+20.27}_{-9.74}$ & $15.99^{+22.31}_{-11.92}$ \\
        Fe & $10.27^{+13.19}_{-5.68}$ & $10.70^{+22.63}_{-5.64}$ & $8.03^{+35.58}_{-5.49}$ & $6.49^{+10.11}_{-3.26}$ & $1.92^{+2.64}_{-1.48}$ & $17.01^{+17.81}_{-7.26}$ & $7.07^{+8.48}_{-4.51}$ \\
        $\tau$($10^{10}$  $\rm cm^{-3}$ s) & $3.01^{+0.63}_{-0.48}$ & $3.17^{+1.26}_{-0.67}$ & $3.25^{+0.76}_{-0.66}$ & $2.14^{+0.39}_{-0.25}$ & $6.11^{+12.27}_{-3.10}$ & $4.47^{+1.65}_{-1.02}$ & $2.54^{+1.84}_{-0.67}$\\
        Normalization ($10^{-4}$${\rm cm}^{-5})$ & $17.0^{+5.0}_{-4.3}$ & $3.2^{+1.4}_{-1.0}$ & $5.4^{+2.9}_{-1.7}$ & $1.9^{+1.0}_{-0.6}$ & $7.1^{+9.7}_{-4.4}$ & $1.4^{+0.9}_{-0.7}$ & $2.2^{+2.6}_{-1.1}$ \\
        C-Statistic (d.o.f.) & 277.33 (172) & 178.67 (151) & 200.76 (154) & 178.03 (153) & 134.62 (138) & 127.53 (130) & 160.86 (160)\\
        \noalign{\smallskip}\hline\noalign{\smallskip}
        ${{n}_{\rm{H}}}$ (${\rm{cm}^{-3}}$) & $0.16\pm0.02$ & $0.25\pm0.05$ & $0.32\pm0.09$ & $0.21\pm0.05$ & $0.50\pm0.35$ & $0.69\pm0.22$ & $0.07\pm0.04$\\
        $t_{\rm ion}$ (kyr) & $4.9\pm1.3$ & $3.4\pm1.5$ & $2.7\pm0.9$ & $2.7\pm0.9$ & $3.2\pm6.8$ & $1.7\pm0.8$ & $9.8\pm9.2$\\
	\noalign{\smallskip}\hline
	\end{tabular}
\end{table*}

\section{Discussion}\label{sec:discussion}
\subsection{Plasma and evolution parameters} \label{sec:evolution}

The plasma density and evolution parameters of G352.7$-$0.1 can be derived using the X-ray spectral fit results  (see Table~\ref{tab:regions}). We estimated the mean hydrogen density $\rm{n_{H}}$ of the post-shock gas using the normalization parameter in Xspec, which is defined as:

\begin{center}
\begin {equation}
        \rm{Normalization}=\frac{10^{-14}}{4 \pi \mathit{d}^2} \int \mathit{n}_{\rm{e}} \mathit{n}_{\rm{H}} d \mathit{V}
	\label{eq:1}
\end{equation}
\end{center}
where ${d=10.5}$~kpc is the distance to G352.7$-$0.1 \citep{Zhang2023}, $n_{\rm{e}}$ and $n_{\rm{H}}$ are the electron and hydrogen densities, respectively, and $V$ is the volume of this SNR. For fully ionized plasma with solar or slightly enhanced metal abundances, we have $n_{\rm{e}}=1.2 n_{\rm H}$. The volume of the entire SNR is calculated by assuming a prolate ellipsoid for the red elliptical region defined in Figure~\ref{fig:rgb}.
The averaged $n_{\rm H}$ value in the SNR is estimated as $0.16\pm 0.02~{\rm cm}^{-3}$.
For the smaller regions in Figure \ref{fig:xrays}, we also calculated $n_{\rm{H}}$ using the same method. 
The three-dimensional (3D) geometry of the X-ray emission in small-scale regions requires more investigation, but here we took a simplifying assumption. 
We assumed that regions ``innertop'', ``cloud'', and ``innerleft'' form a torus (doughnut shape) in 3D. The torus has an inner radius of $1\farcm{18}$, an outer radius of $2\farcm{53}$, and an inclination angle of $40^\circ$. The angle is calculated by $\theta = \rm{arccos}(1.92/2.53)$, where the $1\farcm{92}$ is the semi-minor axis of the yellow ellipse shown in the figure \ref{fig:rgb}.
The central region ``center'' is assumed to be a sphere with a radius of $1\farcm{18}$. The volume of the halo is obtained by subtracting the volume of the torus and central sphere from the ellipsoidal volume of the entire SNR. The bright X-ray spot ``spot'' on the eastern side of the SNR is assumed to be spherical. 
For all the regions, the gas is assumed to be uniformly distributed, although some clumpy structures are present in the X-ray image (see Figure~\ref{fig:rgb}).
The estimated $n_{\rm H}$ values of different regions are shown in Table~\ref{tab:regions}, but these values should be taken with caution as they are provided with an assumed geometry. We also got the X-ray emitting mass of 13.6 $M_{\rm{\odot}}$ for global SNR ($M_{\rm{X}}=1.4n_{\rm{H}}m_{\rm{H}}V$).

Using the ionization timescale $\tau$ and the calculated electron density $n_{\rm e}$ of the global X-ray emitting plasma, we estimate the ionization age by $t_{\rm ion}=\tau / n_e=\tau /1.2n_{\rm{H}}=4.9\pm \rm {1.3~kyr}$. The $t_{\rm ion}$ values of small-scale regions are also calculated and list in Table \ref{tab:regions}. 
The ionization ages across the SNR range from 1 to 6 kyr, except for
the ``halo'' and ``cloud'' regions with large uncertainties.
Considering that ionization age reflects the time after shock heating, the values provided through this method represent only the lower limit of the SNR's age. Finally, we take the ionization age of the global spectrum to infer the lower limit of the SNR age ($4.9\pm 1.3$~kyr).

Assume that the X-ray emission comes from the plasma heated by the forward shock, the forward shock velocity can be inferred by the plasma temperature (for ``snr'' region) as: $V_{\rm s}=\left[16 k T /\left(3 \bar{\mu} m_{\mathrm{H}}\right)\right]^{1 / 2} = 1.3 \pm 0.2 \times 10^3$ km s$^{-1}$, with $m_{\mathrm{H}}$ the atomic mass of hydrogen and the mean atomic weight $\bar{\mu}$ is 0.61 for fully ionized plasma. We found that, apart from the ``cloud'' region, the temperature variation in other regions is not significant. Given that shock velocity can be calculated solely based on temperature, this implies that the shock is slowed down by the dense gas in the ``cloud'' region, while in other regions the shock maintains a relatively uniform expansion velocity. We adopt the Sedov-Taylor phase for this SNR \citep{1959sedov,1950taylor}, since the plasma temperature is moderately high and the gas mass of $13.6~M_\odot$ is larger than typical supernova ejecta mass ($\lesssim 10~M_\odot$). Taking an SNR radius $R_{\rm s}=9.75$~pc (the semi-major axis), we calculated the Sedov age $t_{\mathrm{sedov}}=0.4 R_{\rm s}/V_{\rm s}=2.9 \pm 0.5 $ kyr, smaller than the age obtained from the ionization timescale. 

We obtained a small explosion energy of the SNR in the Sedov phase  \citep{1988ostriker} 
$E_0=25\left(1.4 n_0 m_{\mathrm{H}}\right) R_{\mathrm{s}}^3 V_{\mathrm{s}}^2/(4 \xi) = \left( 1.37 \pm 0.50 \right) \times 10^{50} (n_0/0.04~{\rm cm}^{-3})$~erg.
Here, the ambient density $n_0=n_{\rm H}/4$ is estimated using the average density of the entire SNR  and a compression ratio of 4 for the strong shock. The $\xi$ is a numerical constant with $\xi = 2.026$ for $\gamma=5/3$.
It should be noted that the obtained $E_0$ depends on the SNR radius and the density of the X-ray-emitting gas $n_{\rm H}$. 
Although we find a range of density 0.07--0.7~cm$^{-3}$ in the remnant, the average density in the SNR is small ($n_{\rm H}\approx 0.16$~cm$^{-3}$), while the halo region has the most rarefied medium ($n_{\rm H} \sim 0.07$~cm$^{-3}$).
The explosion energy of G352.7$-$0.1 is likely an order of magnitude smaller than the canonical value of SN ($10^{51}$~erg), hinting at a weak SN explosion.
This is consistent with that obtained by \cite{Kinugasa1998} and \cite{Giacani2009}. \cite{Pannuti2014} got much larger $E_0\sim 2.5\times 10^{51}$~erg by using two-temperature model (0.24 keV and 3.2~keV). 

Since the spectra of G352.7$-$0.1 can be well fit using a single-temperature model, we cannot distinguish whether X-ray-emitting plasma comes from forward shock, reverse shock, or a combination of both. Previous calculations of the SN explosion energy and Sedov age are based on the assumption of forward-shock-heated plasma. If the observed X-ray emission comes from the reverse shock, the ionization age 
% of $4.9\pm 1.3$~kyr 
is still useful to infer the SNR age, and the element abundance ratios and elemental masses are unaffected.

\subsection{Comparison with SN Nucleosynthesis Models}\label{sec:Nucleosynthesis}

\begin{figure*}
	\centering
	\subfloat[CC models (abundance)]{\includegraphics[width=\columnwidth]{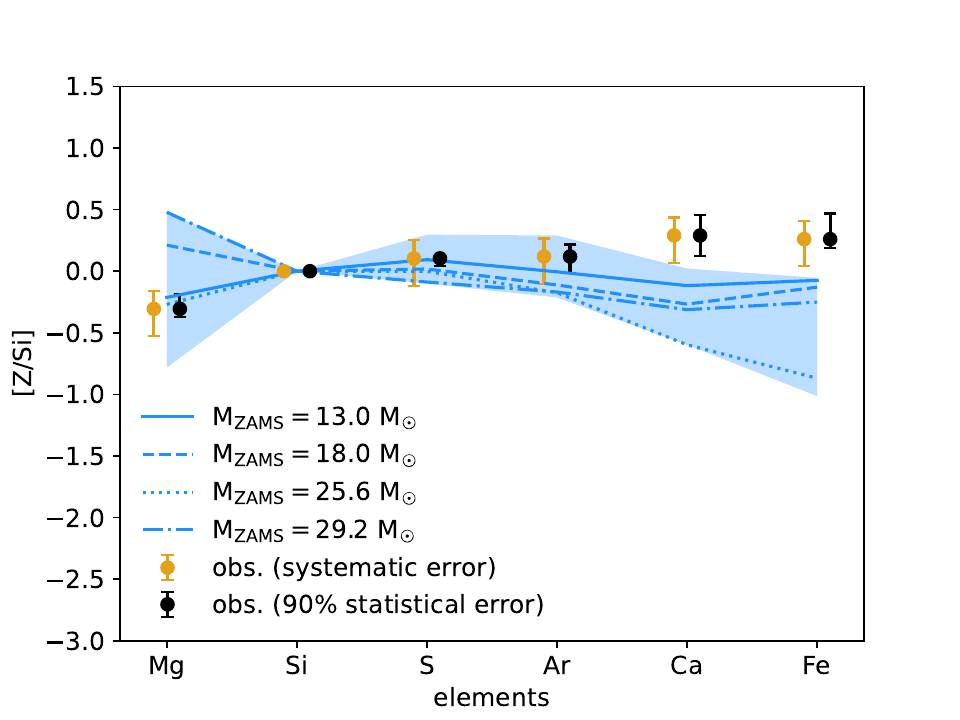}
		\label{compare:a}}
	\subfloat[CC models (mass)]{\includegraphics[width=\columnwidth]{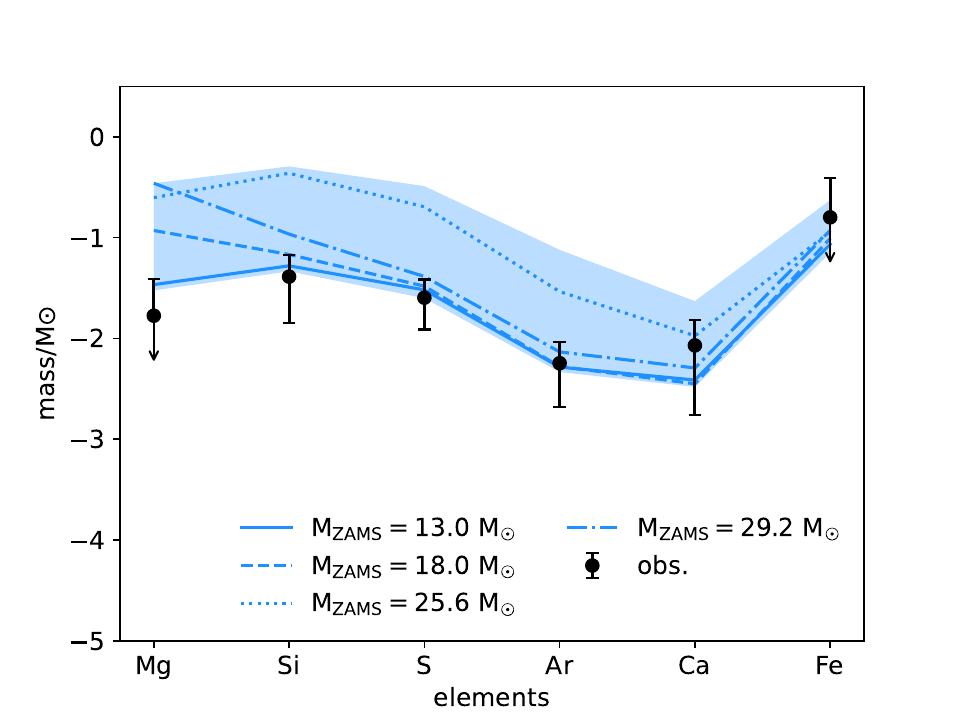}
		\label{compare:b}}
	\caption{Logarithmic abundance ratios (relative to Si) and mass comparison between the observation and core-collapse(CC) nucleosynthesis models. (a) CC SN models for stars with zero-age main-sequence masses from 9 to 120 $M_\odot$\citep{Sukhbold2016}.(b) Consistent with the model in Figure a but compared to mass. We consider all available models in the blue-shaded region, and some of these models are represented with lines.}
	\label{fig:comparecc}
\end{figure*}

\begin{figure*}
	\centering
	\subfloat[DDT models (abundance)]{\includegraphics[width=\columnwidth]{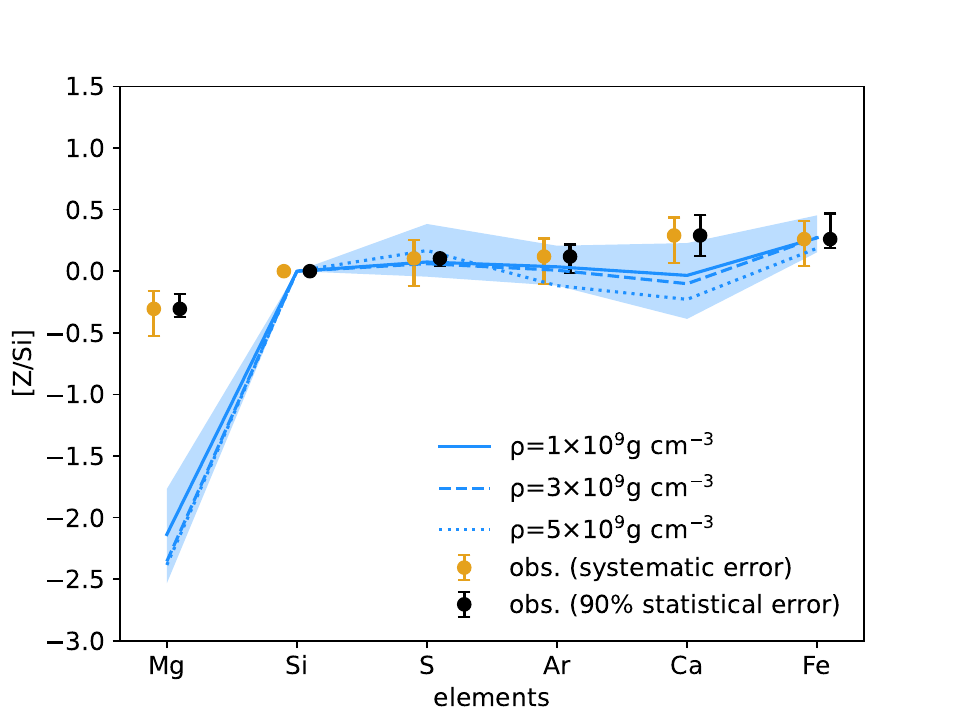}
		\label{compare:c}}
	\subfloat[DDT models (mass)]{\includegraphics[width=\columnwidth]{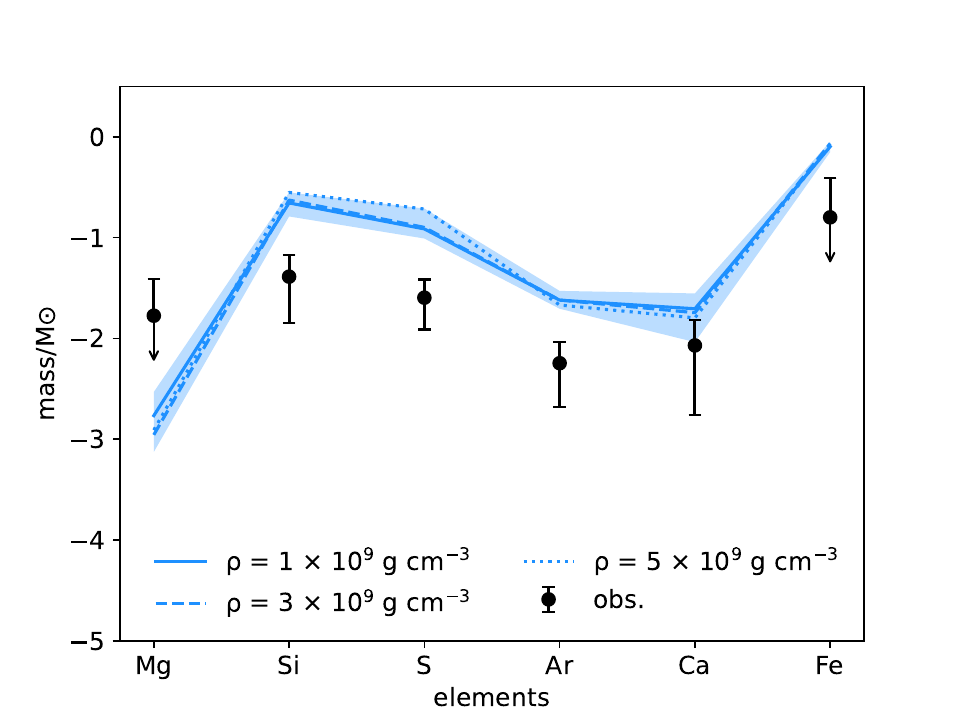}
		\label{compare:d}}
    \hfil
    \subfloat[DD models (abundance)]{\includegraphics[width=\columnwidth]{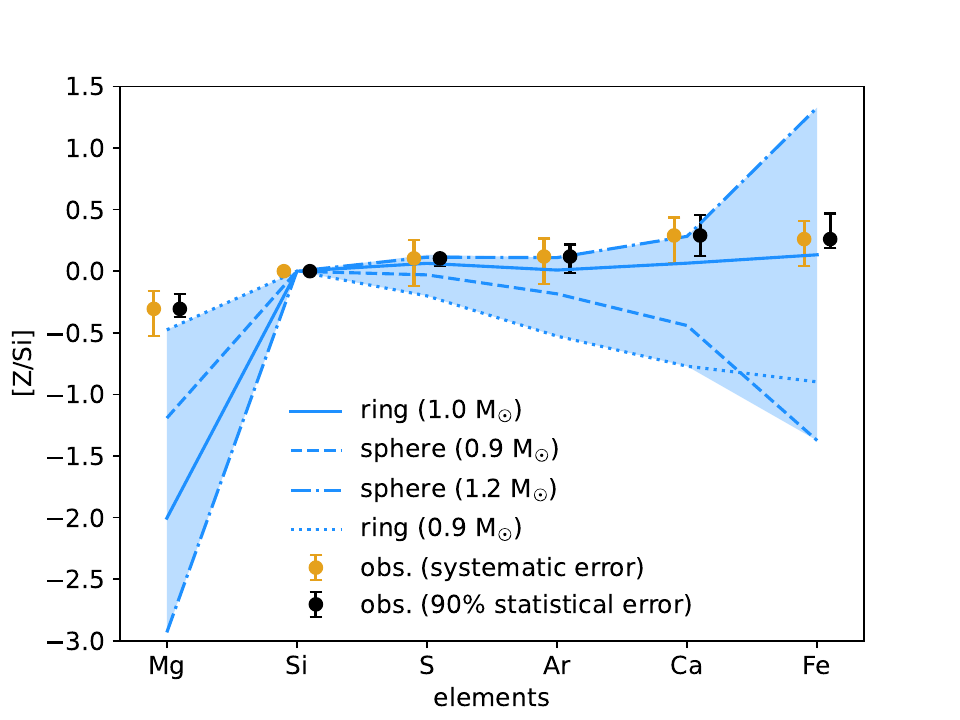}
		\label{compare:e}}
	\subfloat[DD models (mass)]{\includegraphics[width=\columnwidth]{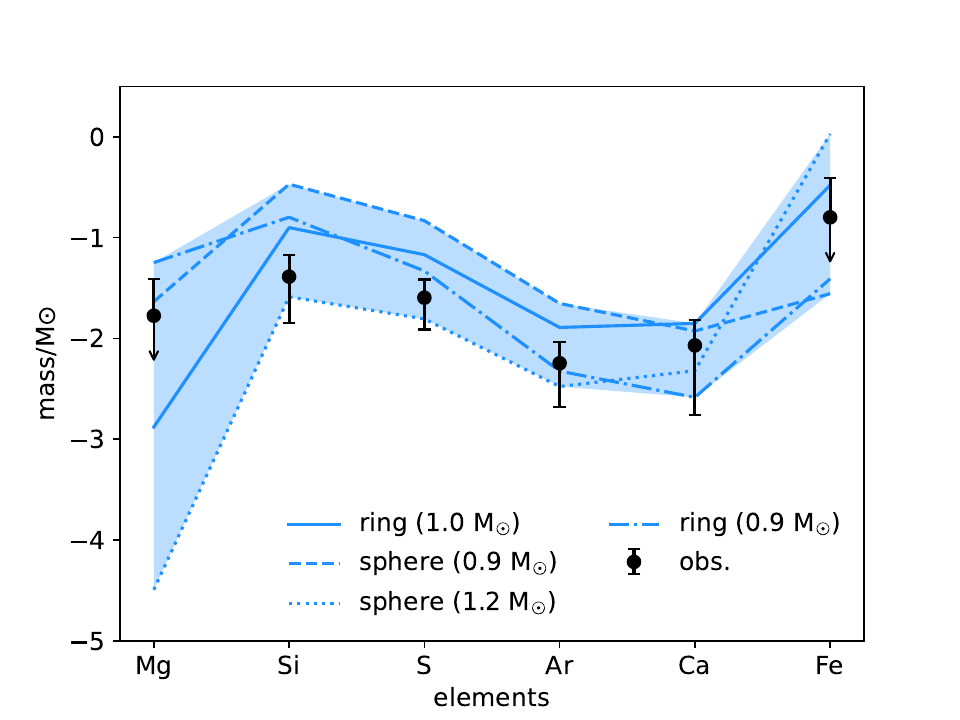}
		\label{compare:f}}
    \hfil
    \subfloat[PTD models (abundance)]{\includegraphics[width=\columnwidth]{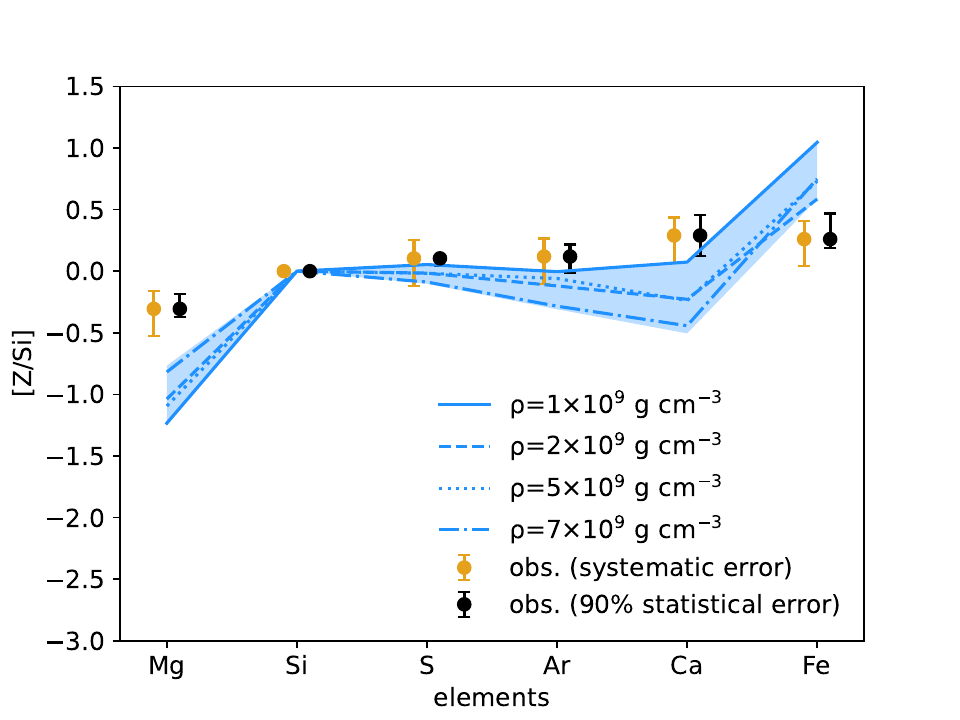}
		\label{compare:g}}
	\subfloat[PTD models (mass)]{\includegraphics[width=\columnwidth]{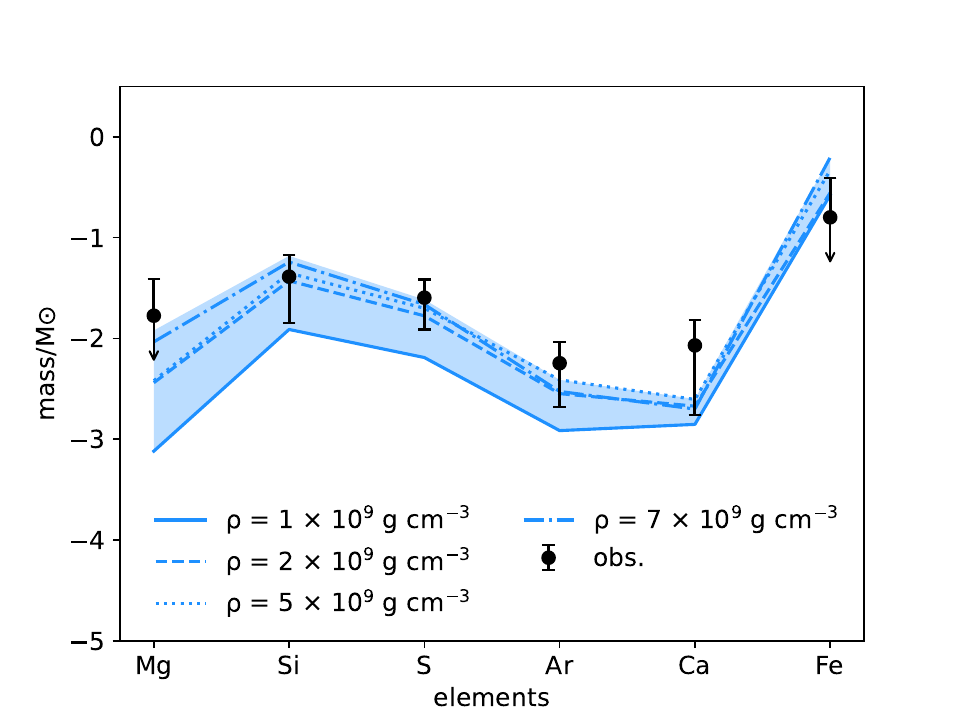}
		\label{compare:h}}  
	\caption{Logarithmic abundance (relative to Si) ratios and mass comparison between the observation and thermonuclear nucleosynthesis models. (a) DDT SN models for white dwarfs with different central densities \citep{leungddt}.(b) Consistent with the model in Figure a but compared to mass. (c) DD Type \uppercase\expandafter{\romannumeral1}a models for sub-Chandrasekhar-mass WDs with different detonation configuration \citep{leungdd2020}. (d)  Consistent with the model in Figure c but compare to mass. (e) PTD Type \uppercase\expandafter{\romannumeral1}a SN models for CO WDs \citep{2020leung1ax}. (f)  Consistent with the model in Figure e but compared to mass. We consider all available models in the blue-shaded region, and some of these models are represented with lines.}
	\label{fig:compare1a}
\end{figure*}

To investigate the SN type and the progenitor of G352.7$-$0.1, we compared its logarithmic abundance ratios with predictions from various SN nucleosynthesis models. 
The logarithmic abundance of element A relative to Si is defined as
$[\mathrm{A} / \mathrm{Si}]=\log _{10}\left(Z_{\mathrm{A}} / Z_{\mathrm{Si}}\right)$, 
where $Z_{\rm{A}}$ is the 
abundance ratio of element A relative to its solar value.
Since the abundances in the remnant are large (3--11$\times$ solar), the observed metals should be dominated by SN ejecta rather than the ISM.
Two groups of uncertainties are provided for the abundance ratios.
Besides the statistical uncertainties obtained from the MCMC method (see Section~\ref{sec:mcmc}), we adopted a systematic error of 40$\%$ for all abundance ratios. 
The systematic uncertainties are introduced by the atomic data \citep{2018Hitomi}, model selections,
as well as 15--40$\%$ biases in the chemical composition determined by the CCD spectra from XMM-Newton or Suzaku \citep{Simionescu_2018}. 
In the logarithmic abundance ratio figures (Figures~\ref{fig:comparecc} and \ref{fig:compare1a}), we show both the 90$\%$ statistic error from MCMC and the systematic error.

Besides the abundance ratios, we also derived the ejecta metal masses and ISM masses as shown in Table~\ref{masstable}, where the metal masses and ISM masses are calculated as 
$M_{\rm A}^{\rm metal}=(Z_{\rm A}-1) M_{\rm H} Z_{\rm A}^\odot/Z_{\rm H}^\odot$ and
$M_{\rm A}^{\rm ISM}= M_{\rm H} Z_{\rm A}^\odot/Z_{\rm H}^\odot$ respectively.
It should be noted that the mass errors are statistical errors, although there should be systematic errors from the assumed morphology of the X-ray-emitting gas.

\begin{table}
	\centering
	\caption{The mass of different elements in ejecta and ISM with 90$\%$ uncertainties}
	\label{masstable}  
	\begin{tabular}{lcc}
	\hline\hline\noalign{\smallskip}	
	Element & ejecta & ISM \\
        &  ($\rm{M_\odot}$) & ($\rm{M_\odot}$) \\
	\noalign{\smallskip}\hline\noalign{\smallskip}
Mg & $<0.039$ & $0.0094$ $\pm$ $0.0014$ \\
Si & $0.041$ $\pm$ $0.027$ & $0.0088$ $\pm$ $0.0013$ \\
S & $0.025$ $\pm$ $0.013$ & $0.0041$ $\pm$ $6 \times 10^{-4}$ \\
Ar & $5.7 \times 10^{-3}$ $\pm$ $3 \times 10^{-3}$ & $8 \times 10^{-4}$ $\pm$ $1 \times 10^{-4}$ \\
Ca & $8.5 \times 10^{-3}$ $\pm$ $7 \times 10^{-3}$ & $9 \times 10^{-4}$ $\pm$ $1 \times 10^{-4}$ \\
Fe & $<0.39$ & $0.017$ $\pm$ $3 \times 10^{-3}$ \\
    
	\noalign{\smallskip}\hline
	\end{tabular}
\end{table}

\subsubsection{Core-collapse Models}

 \cite{Sukhbold2016} 
provided SN nucleosynthesis products for massive stars with solar metallicity and zero-age main-sequence masses in the 9.0 -- 120 $M_{\odot}$ range. 
We used the yields from  the N20 central engine \footnote{We also tested the W18 central engine \citep{Sukhbold2016} and obtained similar results. The best fit for the progenitor star mass was 12.25 $\rm{M_\odot}$ ($\chi^2$ = 4.73).}
\citep{Sukhbold2016,NOMOTO198813} and took the metals of the progenitor winds into account.  Figure \ref{fig:comparecc} shows the comparison between the observation and the core-collapse SN models. 
The observed masses of various elements can be well fitted by the nucleosynthesis model of a core-collapse SN with a progenitor mass of 13 $\rm{M_{\odot}}$. 
The logarithmic abundance ratios [Mg/Si], [S/Si], and [Ar/Si] are consistent with the 13~$M_\odot$ core-collapse SN model, but the observed [Ca/Si] and [Fe/Si] are larger than model prediction ($\chi^2$ = 4.86 for abundance ratios and using the systematic error).

\subsubsection{Type \uppercase\expandafter{\romannumeral1}a Models}

Type Ia SNe result from thermonuclear explosions of white dwarfs (WDs).
However, there is currently no consensus on their progenitor or explosion mechanism \citep[e.g., see the most recent review and references therein,][]{liu2023type}. The delayed detonation (DDT) model and the double detonation (DD) model are currently popular explosion mechanisms for Type \uppercase\expandafter{\romannumeral1}a SNe \citep{Maeda2022}.
In recent years, more explosion models, such as pure turbulent deflagration models and He-shell detonation models, have also been proposed for explaining peculiar thermonuclear SNe \citep{Fink_2013,2020leung1ax,2011Waldman}.

In the DDT model, the  Chandrasekhar mass ($\rm{M_{Ch}}$) WD explosion starts from a subsonic deflagration and then transfers to a detonation.
DDT models are found to well explain light curves and spectral evolution of normal Type \uppercase\expandafter{\romannumeral1}a SNe \citep[e.g.,][]{Maeda2016}). 
In this paper, we compare the observed results with the DDT nucleosynthesis results from \cite{leungddt}, in which they used a 2D hydrodynamic model to calculate the SN nucleosynthesis products.
Figures \ref{compare:c} and \ref{compare:d} show the observed values and the DDT models with different WD central densities. The comparison shows a significant difference between the observed abundance ratio [Mg/Si] and that predicted by the DDT models (minimum $\rm{\chi^2}$ = 6.96). 
The observed masses of metal elements also largely deviate from the models. This indicates that the DDT model cannot adequately describe SNR G352.7$-$0.1.

\begin{figure*}
	\centering
	\subfloat{\includegraphics[width=\columnwidth]{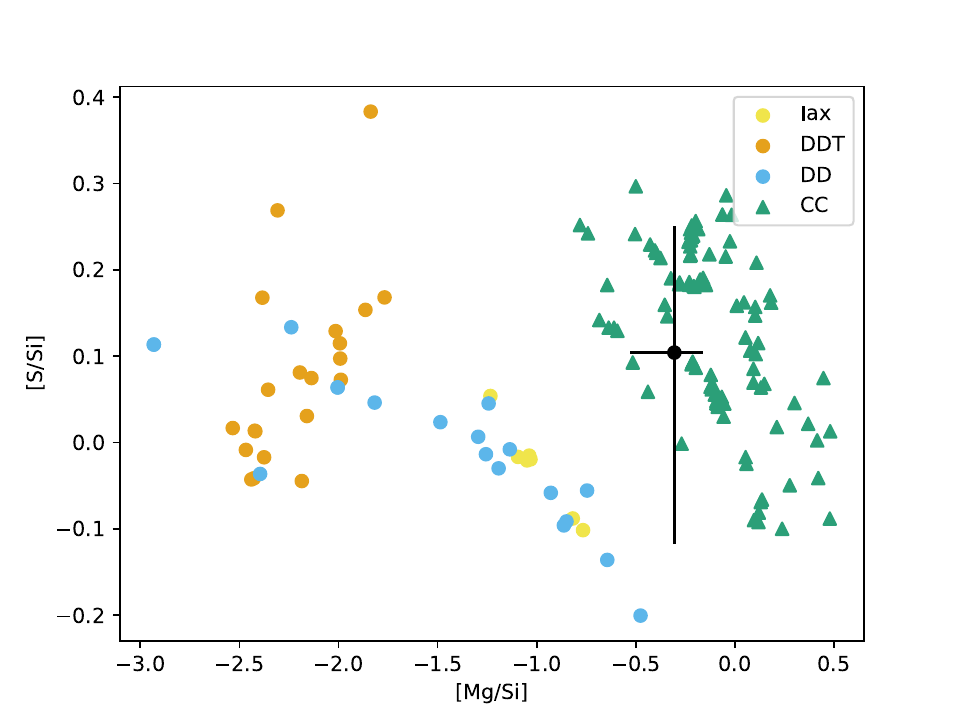}
		\label{[Mg/Si]-[S/Si]diagram}}
	\subfloat{\includegraphics[width=\columnwidth]{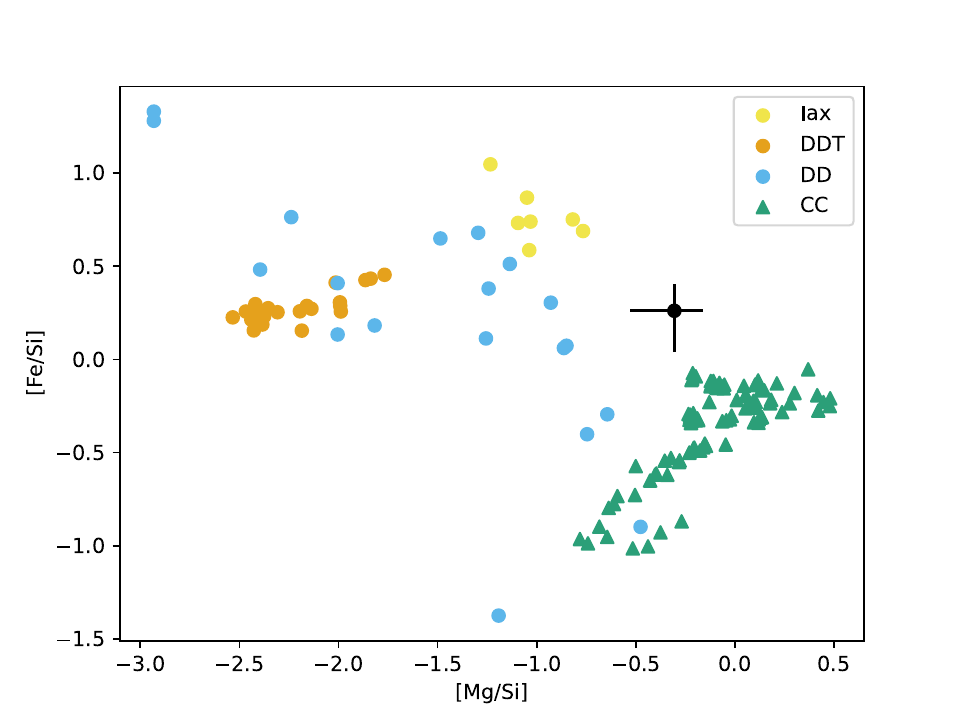}
		\label{[Mg/Si]-[Fe/Si]diagram}}
	\caption{[Mg/Si]–[S/Si] diagram and [Mg/Si]-[Fe/Si] diagram for a comparison between SNR G352.7$-$0.1 and four SN nucleosynthesis models (see also Figures~\ref{fig:comparecc} and \ref{fig:compare1a}).}
	\label{fig:comparemgsi}
\end{figure*}

In addition to the DDT mechanism, the DD mechanism, as a sub-$\rm{M_{Ch}}$ WD explosion model, can also explain some properties of Type \uppercase\expandafter{\romannumeral1}a SNe.
We took the DD nucleosynthesis models from \cite{leungdd2020}, which provided a few benchmark models for describing normal Type Ia SNe and  
also, explored a large parameter space for studying Type Ia SN diversity (peculiar Type Ia SNe).
Under this DD explosion mechanism, the surface of a sub-$\rm{M_{Ch}}$ WD first undergoes a He detonation. This detonation is not powerful enough to explode the entire WD, but when the detonation wave propagates, it can trigger a carbon detonation inside the WD, leading to the production of a Type \uppercase\expandafter{\romannumeral1}a SN. 
Figures \ref{compare:e} and \ref{compare:f} compare these DD models with our observations of G352.7$-0.1$. 
The abundance ratio plot shows that the models either underpredict the [Mg/Si] value or the [S/Si]--[Fe/Si] values (minimum $\rm{\chi^2}$ = 7.79 for DD models).
In the elemental mass comparison, none of the DD models can satisfactorily fit the observed mass of Mg and other elements simultaneously.

Finally, we considered the pure turbulent deflagration (PTD) models of near-$\rm{M_{Ch}}$ WDs, which are proposed to explain the subclasses of Type Ia SNe, especially for those sublumionous ones \citep{Fink_2013,2020leung1ax}\footnote{There are also models for peculiar sub-luminous thermonuclear SNe called Ca-rich transients and have been used for SNRs \citep[see][and references therein]{Weng_2022}, but we do not discuss these models because the Ca abundance in G352.7$-0.1$ is not extremely high compared to other elements.}
Unlike the DDT explosion mechanism, the flame in PTD models propagates subsonically and does not trigger a detonation at the later stages.
The pure deflagration cannot completely unbind the WD but is quenched as the WD expands, leaving unburnt materials.
This mechanism can cause a weak SN explosion compared to that of a normal Type Ia SNe, so that might describe the low explosion energy of G352.7$-0.1$.
The PTD models used in this paper are adopted from \cite{2020leung1ax}, which explained the metal pattern in Sgr~A~East \citep{zhousgr2021}. 
As shown in Figures \ref{compare:g} and \ref{compare:h}, all the models predict too-low [Mg/Si] and too-large [Fe/Si] ratios compared to those of the SNR  (minimum $\rm{\chi^2}$ = 16.43), although several models can fit the observed metal masses.

The above comparisons show that none of the SN models perfectly describe all the metal abundance ratios, however, CCSN models with a progenitor mass of $\sim 13~M_\odot$ provide a relatively good fit with $\chi^2=4.86$.
In general, the CCSN models underpredict the [Ca/Si] and [Fe/Si] ratios in G352.7$-0.1$, while Type Ia SN models fail to reproduce the high [Mg/Si] ratio.
Not all models are shown with lines in  Figure~\ref{fig:comparecc} and Figure~\ref{fig:compare1a}. As a supplement, we plot all the models with their [Mg/Si], [S/Si], and [Fe/Si] ratios in Figure~\ref{fig:comparemgsi}, in comparison with observed values and the $40\%$ uncertainties.
The [Mg/Si] and [S/Si] ratios prefer a CC origin of G352.7$-0.1$, while the [Mg/Si] and [Fe/Si] ratios cannot be explained with the CC or Ia models. 
It is noteworthy that Ca and Fe abundances have the largest statistical uncertainties using our single thermal plasma model (see Table~\ref{tab:regions}), and the two-thermal component model using Suzaku data by \cite{Sezer2014} provided a different Fe abundance.
Therefore, the abundance ratios of Mg, Si, S, and Ar to Si are better constrained than Ca/Si and Fe/Si.

\subsection{Further discussion about the SN type}

Based on the chemical composition, we suggest a CC origin for G352.7$-$0.1. 
On the contrary, this SNR was proposed to be a Type Ia SNR, since the Fe K$\alpha$ line centroid ($<6.55$~keV; determined by the ionization state) falls in the range for a Type Ia SNR evolving a uniform interstellar medium with a density between $\rm{0.6-3\ cm^{-3}}$ \citep{Yamaguchi2014,Sezer2014}.
The ionization state of NEI plasma highly depends on the environmental density in which it impacts, with a lower K$\alpha$ line centroid energy corresponding to a less dense gas. 
SNR G352.7$-0.1$ expands in an inhomogeneous medium, and the X-ray plasma density is between 0.07 and 0.5~cm$^{-3}$ (see Table \ref{tab:regions}). The estimated average density of the ambient medium is only 0.04~cm$^{-3}$ (see Section \ref{sec:evolution}), which is two orders of magnitude less than that used for the Fe K$\alpha$ line centroid method. 
Therefore, the low energy of the Fe K$\alpha$ line centroid from G352.7$-0.1$ is consistent with an SNR evolving in a low-density medium, but should not be regarded as firm evidence of a Type Ia origin \citep[see also discussions in][]{Siegel2021}.
Another implication of CC origin comes from the previous molecular studies. 
\cite{Zhang2023} found an expanding molecular bubble surrounding the SNR and the wind bubble size implies a $\sim 12 M_\odot$ progenitor star.
This is consistent with our suggestion based on the chemical composition.

\subsection{Explanation of the mixed morphology}\label{sec:4.3}

Standard Sedov evolution of an SNR in a uniform medium predicts a shell-like X-ray morphology, but the X-ray emission of G352.7$-0.1$ is centrally filled.
Several possible mechanisms have been used to explain the centrally filled X-ray morphology of MMSNRs, such as thermal conduction \citep{Cox1999ApJ...524..179C, Shelton1999ApJ...524..192S}, the effect of evaporating clouds \citep{White1991ApJ...373..543W}, shock reflection \citep{Chen2008ApJ...676.1040C, Zhang2015ApJ...799..103Z}, and projection effects \citep{Petruk2001A&A...371..267P, Zhou2016ApJ...831..192Z, Zhang2019sros.confE..30Z}. Below we discuss the possibilities of these models in G352.7$-0.1$.

Thermal conduction: 
Efficient thermal conduction may lead to smooth and near-uniform distributions of the temperature and density in the SNR interior 
\citep{Cox1999ApJ...524..179C,Shelton1999ApJ...524..192S}. This allows an increase of the density in the SNR center and thus an enhancement of the X-ray emission. If we ignore the suppression of the magnetic field, the thermal conduction timescale is estimated as $t_{\text {cond }} \approx k n_e l_T^2 / \kappa \sim 56\left(n_e / 1 \mathrm{~cm}^{-3}\right)\left(l_T / 10 \mathrm{pc}\right)^{2}(k T / 0.6 \mathrm{keV})^{-5 / 2}(\ln \Lambda / 32) \mathrm{kyr}$, where $l_T$ is the length scale of the temperature gradient, $\kappa$ is the collisional conductivity, 
and the Coulomb logarithm is $\ln \Lambda = 29.7 + \ln \left( n_e^{-1/2} \left( {T_{\rm{X}}}/{10^6 \text{ K}} \right) \right)$ \citep{spitzer1962physics}. 
Taking ${l_T} = \rm{7.73\ pc}$ from the short semi-axis of the SNR, 
a temperature of $kT_{\rm X}$ = 2.04 keV 
and an electron density of $n_e=0.33\ \rm{cm^{-3}}$ for the X-ray bright region in the yellow ellipse as shown in Figure \ref{fig:rgb}, we obtained the thermal conduction timescale of $\sim 0.54$ kyr.
The presence of magnetic fields can significantly suppress thermal conduction. 
However, if we consider the presence of chaotic magnetic field fluctuations, compared to an ordered field, thermal conduction still works at some level
\citep{1978Rechester}.
In this case, the collisional conductivity $\kappa$ could be reduced to $\gtrsim 1/5$ the classical value \citep{Lazarian_2006, 2001Narayan}. Consequently, the thermal conduction timescale for plasma with chaotic magnetic fields in the SNR would be below around 2.7 kyr, which is still smaller than the SNR age.
This indicates that thermal conduction likely has some influence on the SNR.
However, the thermal conduction process alone is insufficient to explain all the X-ray properties in G352.7$-0.1$. 
The overall X-ray brightness of the SNR is not smooth, but much concentrates inside the inner radio ring. The gas density abruptly drops from the inner ring to the halo region, requiring an extra process to explain the large density gradient.

Evaporating clouds: 
In the model proposed by \cite{White1991ApJ...373..543W}, the SNR shock propagates through a cloudy medium in the SNR center and some dense cloudlets survive. These cloudlets slowly evaporate in the hot medium, increasing the density of the interior hot gas and diluting the ejecta to some extent. 
Considering that G352.7$-0.1$ is impacting molecular clouds in the southern part of the inner shell \citep{Zhang2023}, some evaporated clouds may contribute to the X-ray emission at the interaction region (``cloud'') at the SNR boundary.
Nevertheless, the evaporating cloud mechanism still has some difficulties, especially in explaining the chemical properties of the SNR interior. 
If the central X-ray emission was produced by evaporating clouds, the central region (``inner'') should show significantly low metal abundances, which is clearly inconsistent with observations (see Table~\ref{tab:regions}).

Shock reflection: 
Shock reflection has been used to explain some morphology of MMSNRs, such as Kes 27 and G337.8$-$0.1 \citep{Zhang2015ApJ...799..103Z,Chen2008ApJ...676.1040C}. 
This mechanism considers that the interaction of the shock with a dense cavity wall generates a fast shock reflected to the SNR center, and a slow transmitted shock propagating into the cavity wall.
It predicts a cold SNR periphery and a hot interior.
Although G352.7$-0.1$ is proposed to be evolving in a low-density cavity created by its progenitor winds \citep[see also][]{Giacani2009,Zhang2023},
our analysis does not reveal a significantly higher temperature in the inner region (``center'') compared to the outer part.
The molecular observation does not reveal a complete cavity wall surrounding the remnant (see Figure~\ref{fig:rgb}), but 
there only exists a high-density cloud in the inner ring.

Projection effect: 
The brightest X-ray emission of SNR G352.7$-$0.1 well correlates with its inner radio ring, as shown in Figure \ref{fig:rgb}.
The shell-like radio emission often traces the boundary of the SNR, indicating that the inner radio ring is a peripheric structure projected in the SNR interior.
Besides the inner ring, the radio structures of G352.7$-0.1$ contain an NE shell and an SW shell. The radio morphology may match a ``barrel-shaped'' model with a peculiar viewing angle as proposed by \cite{1987Manchester} and \cite{Giacani2009}.
Another possible 3D geometry of the SNR is an hourglass-shape. 
Viewing at a certain angle, the SNR's 2D shape could consist of the inner ring and bipolar shells (e.g., SN~1987A).
The formation of the hourglass-shaped morphology can be attributed to the bipolar progenitor wind bubbles or an interaction of the SNR with the dense interstellar medium in the inner radio ring region. 
Compared to the barrel-shaped morphology, the hourglass-shaped morphology better explains the correlation between the bright X-ray emission and the inner ring, where the density is enhanced due to either the circumstellar medium at the equatorial ring or the denser interstellar medium.
Therefore, we suggest that the central X-ray emission results from a projection effect, although other processes, such as thermal conduction, may play a role.
This mechanism was previously used to explain MMSNR Kes~79 \citep{Zhou2016ApJ...831..192Z}, which also reveals a multi-shell morphology in the radio band.

In addition, the rarefaction scenario has also been used to explain MMSNRs. It assumes that the SN explodes in a high-density medium, and the plasma heated by the shock quickly reaches ionization equilibrium. When the blast wave breaks out of the interior dense medium, the shocked plasma cools rapidly due to adiabatic expansion, and eventually enters the over-ionized state \citep{1989Itoh, Shimizu_2012}. This scenario has been used to explain some MMSNRs, such as IC 443 \citep{2009Yamaguchi} and W28 \citep{Okon_2018}. In some cases, this rarefaction and thermal conduction between the hot and cold dense medium work together to shape the SNR morphology, explaining SNRs such as W49B \citep{2006Miceli,2011zhouxin}. However, in SNR G352.7$-$0.1, we did not find the presence of over-ionized plasma.

\section{Conclusion}\label{sec:5}

In this paper, we conducted a spatially resolved X-ray spectral analysis of SNR G352.7$-$0.1 using the data from the XMM-Newton X-ray telescope. We also compared the metal pattern of this SNR with various SN nucleosynthesis models to investigate its SN origin. We suggest that CCSN models better explain the abundance pattern and the ejecta mass observed in this SNR. Our conclusions are summarized below:

\begin{enumerate}
    \item 
The XMM-Newton X-ray spectrum of G352.7$-$0.1 shows emission lines of Mg, Si, S, Ar, Ca, and Fe. The 0.8--7 keV spectra can be satisfactorily fit with an absorbed NEI plasma model, 
with a temperature $kT\sim2.1^{+0.7}_{-0.2}$~keV and ionization timescale 
$\tau\sim\rm{3.0^{+0.6}_{-0.5}\ \times 10^{10} \ cm^{-3}\ s}$ respectively.
We obtained the abundances relative to the solar values of Mg ($2.8^{+2.3}_{-1.1}$), Si ($5.6^{+2.9}_{-1.4}$), S ($7.2^{+3.1}_{-1.4}$), Ar ($7.4^{+4.0}_{-2.3}$), Ca ($11.0^{+7.8}_{-4.9}$) and Fe ($10.3^{+13.2}_{-5.7}$). The error is provided at the 90\% confidence level. 

\item 
Our spatially resolved analysis shows that the abundances and temperature do not vary significantly across the SNR, except for the region labeled as ``cloud'', where \cite{Zhang2023} found evidence of an interaction between the SNR and molecular cloud. We observed low metal abundances and temperature, and an enhanced density in this region. 
This further supports the idea that the SNR interacts with dense gas in the southern part of the inner radio ring.
\item 
We employed the MCMC method to calculate the abundance ratios within the 90$\%$ confidence level, and obtained Mg/Si=0.43--0.66, S/Si=1.11--1.35, Ar/Si=0.97--1.65, Ca/S=1.33--2.84, Fe/Si=1.54--2.94.
\item
We obtained the average hydrogen density of the X-ray-emitting plasma of $\sim 0.16~\rm{cm^{-3}}$ and an ionization age of $\sim 5$ kyr. 
\item 
By comparing the observed metal pattern and that predicted by the SN nucleosynthesis models, we proposed that G352.7$-0.1$ likely originated from a CCSN with a progenitor mass of 13 $\rm{M_\odot}$.

\item
We discussed the possible mechanisms that could cause the centrally filled X-ray emission in G352.7$-0.1$.
Due to the good correlation between the X-ray emission and the inner radio ring, the favored mechanism is a projection effect.
\end{enumerate}

\section*{Acknowledgements}
We are grateful to Gloria Dubner for providing the VLA radio images of G352.7$-$0.1.
This article is based on data from the XMM-Newton with the ID 0150220101. We also thank Gao-Yuan Zhang, Jian-Bin Weng, Hiroya Yamaguchi, and Yang Chen for their valuable comments and helpful discussion. This research utilized the Science Analysis System (SAS) for preliminary processing of XMM-Newton data. Further analysis was conducted using software tools from the High Energy Astrophysics Science Archive Research Center (HEASARC). 
This publication is based on data acquired with the Atacama Pathfinder Experiment (APEX) under program ID 0103.D-0387 (A). APEX is a collaboration between the Max-Planck-Institut fur Radioastronomie, the European Southern Observatory, and the Onsala Space Observatory. 
L.-X.D. and P.Z acknowledge the support from the National Natural Science Foundation of China (NSFC), under No.\ 12273010. 
L.S.\ acknowledges the support from Jiangsu Funding Program for Excellent Postdoctoral Talent (2023ZB252).

    \textit{Software:} DS9 \citep{2003Ds9_Joye}, SAS \citep{2014SAS}, Xspec \citep{1996xspec_Arnaud}.

%%%%%%%%%%%%%%%%%%%%%%%%%%%%%%%%%%%%%%%%%%%%%%%%%%
\section*{Data Availability}
 
The data in this article are available in the XMM-Newton Science Archive, at https://www.cosmos.esa.int/web/xmm-newton/xsa/. The final data products will be shared on reasonable request to the authors.

\bibliographystyle{mnras}
\bibliography{example} 

\section*{Appendix}
\section*{X-ray spectra of small-scale regions in SNR G352.7$-$0.1}
We show the spectra and the best-fit models of small-scale regions in Figure~\ref{fig:xrays}. 
The best-fit parameters are shown in Table~\ref{tab:regions}.
\begin{figure*}
	\centering
	\subfloat[center]{\includegraphics[width=\columnwidth]{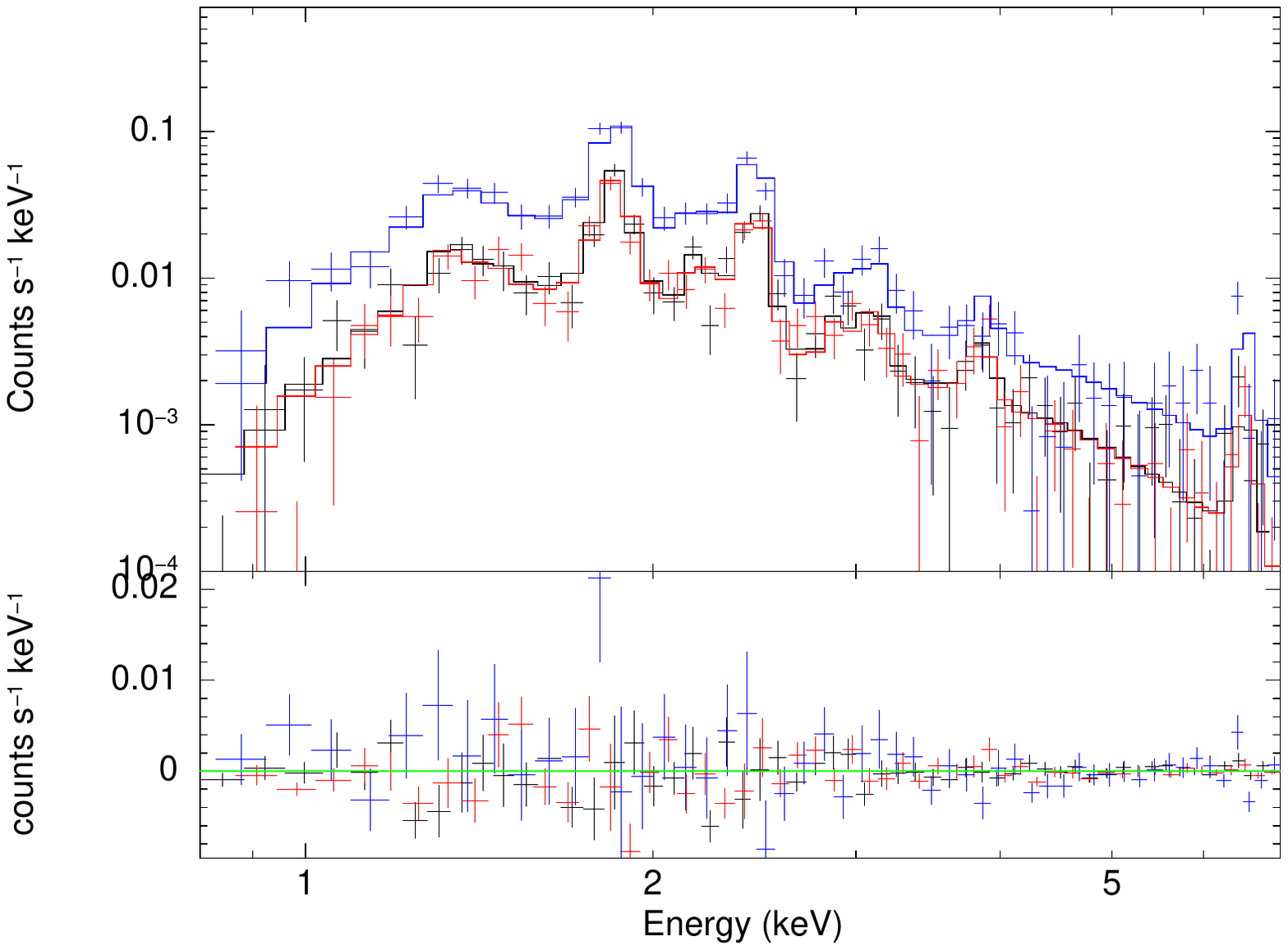}
		\label{spec1}}
	\subfloat[innerleft]{\includegraphics[width=\columnwidth]{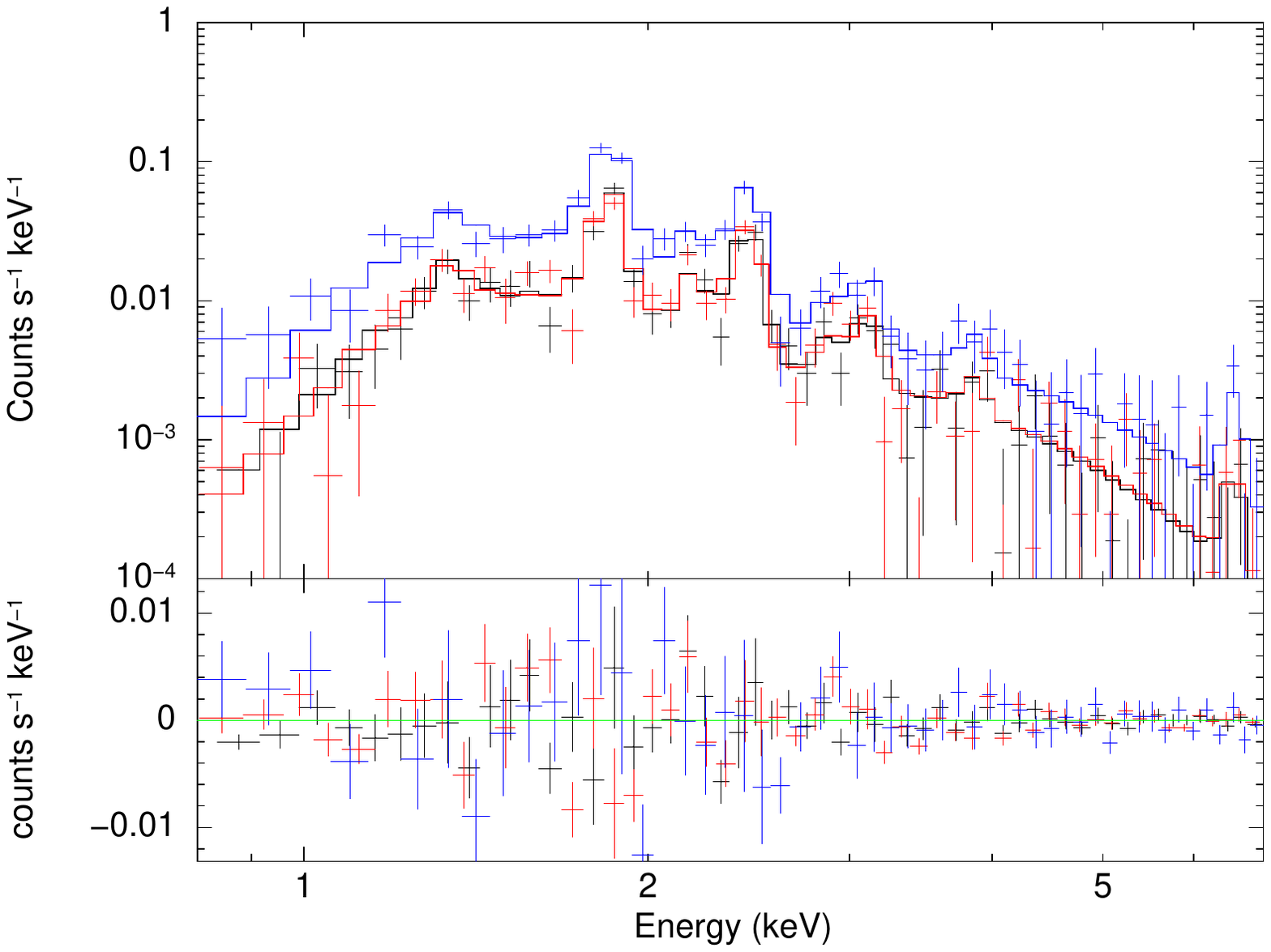}
		\label{spec2}}
    \hfil
    \subfloat[innertop]{\includegraphics[width=\columnwidth]{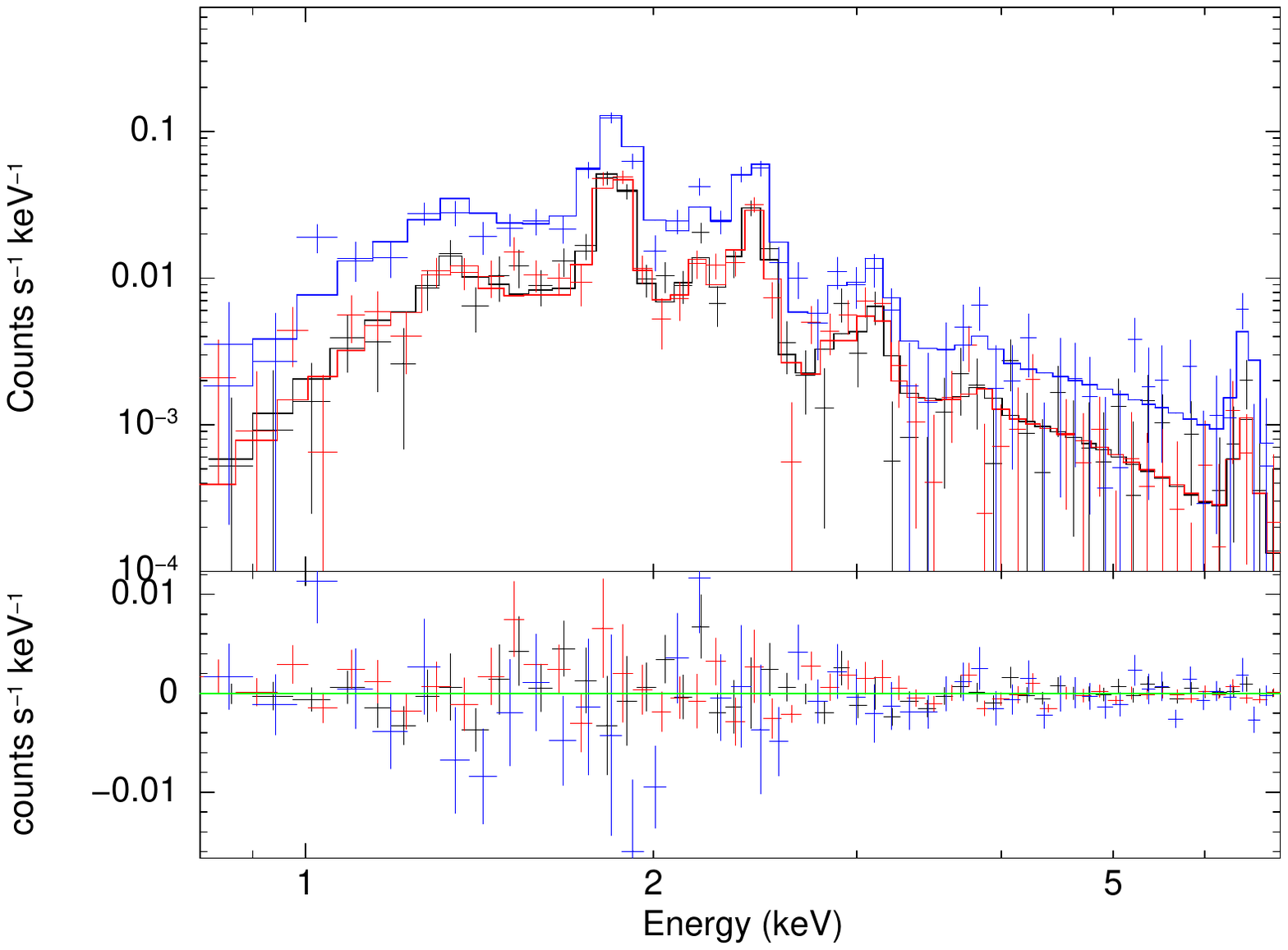}
		\label{spec3}}
	\subfloat[cloud]{\includegraphics[width=\columnwidth]{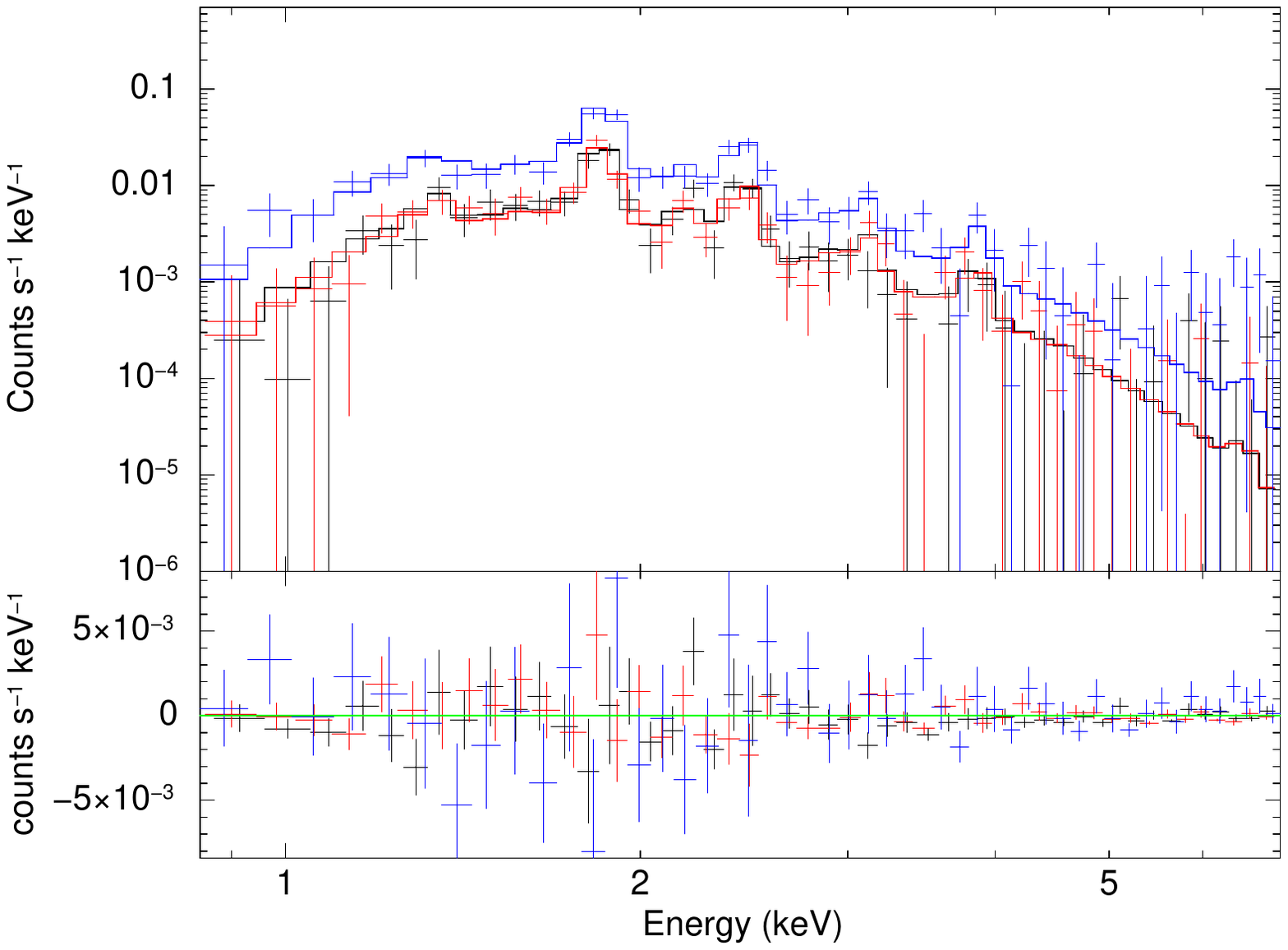}
		\label{spec4}}
    \hfil
    \subfloat[spot]{\includegraphics[width=\columnwidth]{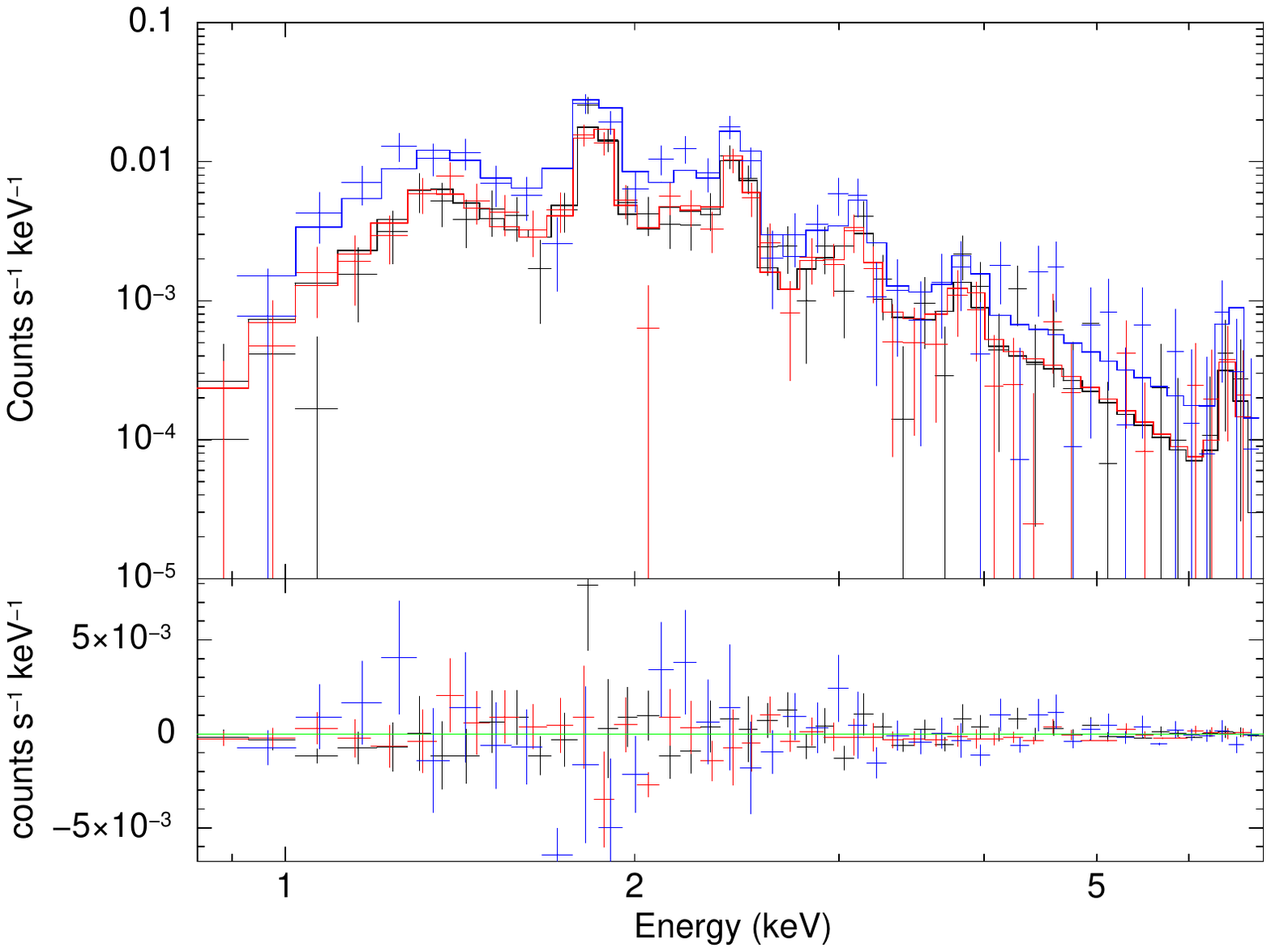}
		\label{spec5}}
	\subfloat[halo]{\includegraphics[width=\columnwidth]{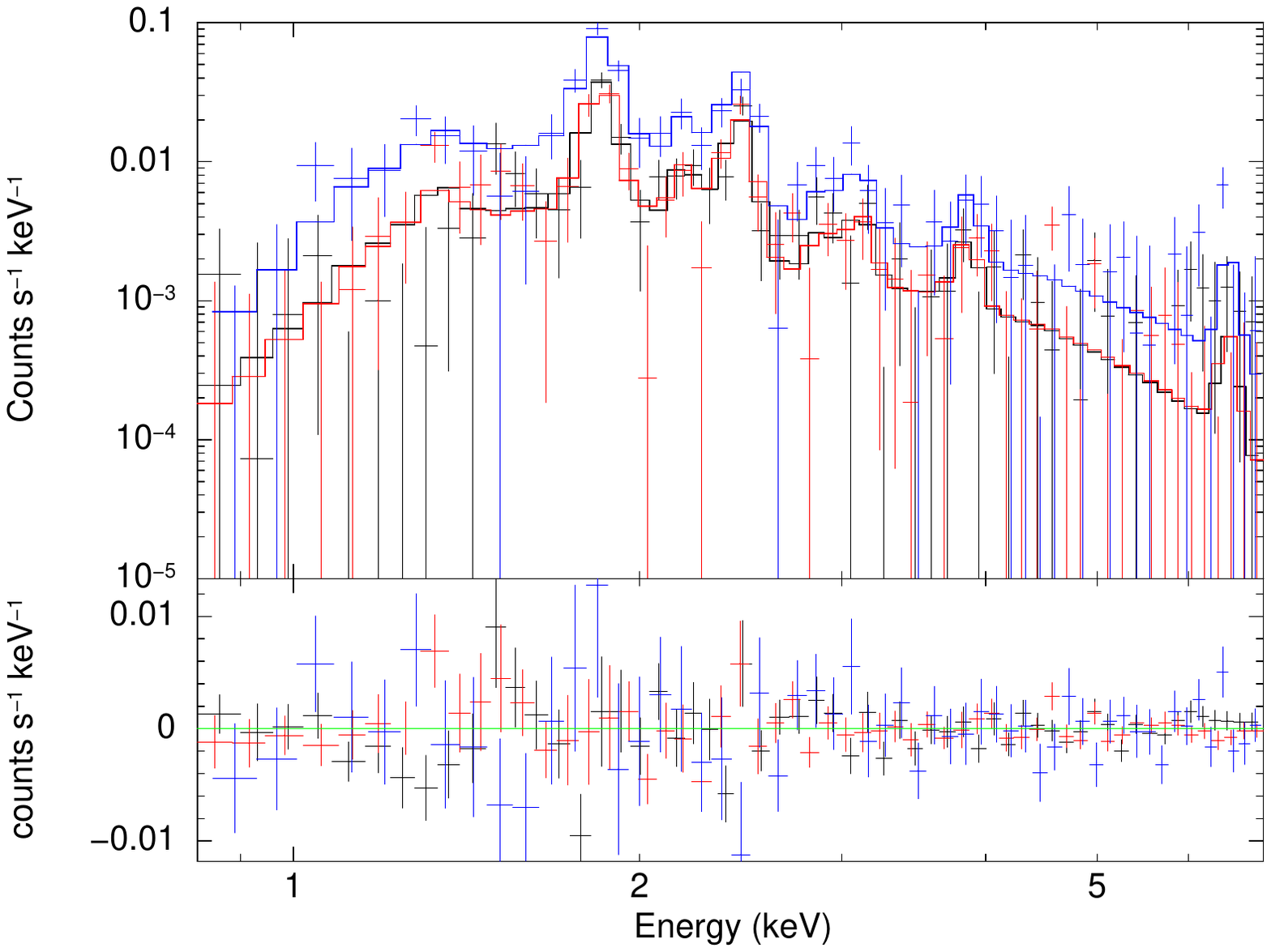}
		\label{spec6}}  
	\caption{XMM-Newton MOS1, MOS2, and pn spectra in 0.8--7 keV (black, red, and blue, respectively) for small scale regions labeled in (see Figure~\ref{fig:xrays}, fitted with absorbed $tbabs$×$vnei$ models. The best-fit results are tabulated in Table~\ref{tab:regions}.}
	\label{fig:spec}
\end{figure*}

\bsp	
\label{lastpage}
\end{document}